\documentclass[fleqn,twoside]{article}       % Publisher version
\usepackage{espcrc2}
\usepackage{subeqn}
\usepackage{hyperref}
\usepackage{axodraw}

\newcommand{\sgn}{\ensuremath{\mathrm{sgn\ }}}
\newcommand{\lsim}{\buildrel < \over{_\sim}}
\newcommand{\gsim}{\buildrel > \over{_\sim}}

\begin{document}
\begin{abstract}
In the first part of the talk, three key ideas proposed in the 1970s,
and in particular their \emph{combined role} in providing an understanding
of the neutrino-masses as well as of the baryon-asymmetry of the universe, are
expounded.  The ideas in question include: (i) The symmetry $SU(4)$-color,
which introduces the right-handed neutrino as an \emph{essential member}
of each family and also provides (rather reliably) the Dirac mass of the
tau-neutrino by relating it to the top quark mass; (ii) SUSY grand
unification together with the scale of the meeting of the three gauge
couplings, which provides the scale for the superheavy Majorana masses of the
RH neutrinos; and (iii) the seesaw mechanism, which combines the Dirac and
the superheavy Majorana masses of the neutrinos obtained as above to yield
naturally light LH neutrinos and in particular the \emph{right magnitude}
for $m(\nu_{L}^{\tau})$.  In the second part, an attempt is made, based in
part on recent works, to show how a set of diverse phenomena including
(a) fermion masses, (b) neutrino oscillations, (c) CP and flavor violations,
and (d) baryogenesis via leptogenesis can fit together neatly within
\emph{a single predictive framework} based on an effective symmetry group
\(G(224) = SU(2)_L \times SU(2)_R \times SU(4)^c\) or $SO(10)$,
possessing supersymmetry.  CP and flavor violations arising within this
framework include enhanced rates (often close to observed limits) for
$\mu \rightarrow e\gamma$ and $\tau \rightarrow \mu \gamma$ and
also measurable electric dipole moments of the neutron and the electron.
Expectations arising within the same framework
for proton decay are summarized at the end.  It is stressed that the
two notable missing pieces of this framework,
which is otherwise so successful, are
supersymmetry and proton decay.  While the search for supersymmetry at the
LHC is eagerly awaited that for proton decay will need the building of a
next-generation megaton-size underground detector.
\end{abstract}
\title{\textbf{Neutrino Masses: Shedding Light on Unification and Our
Origin}\thanks{Invited talk presented at the Fujihara seminar on ``Neutrino
Mass and Seesaw Mechanism'', held at the KEK Laboratory, Tsukuba, Japan
(Feb23-25, 2004)}}
\author{Jogesh C. Pati\\
Department of Physics\\
University of Maryland\\
College Park, MD 20740-4111 USA}
\maketitle
\setcounter{footnote}{0}

\section{Introduction}
Since the discoveries (confirmations) of the atmospheric \cite{sk}
and solar neutrino oscillations \cite{sno,Bahcall}, the neutrinos have
emerged as being among the most effective probes into the nature of
higher unification.  Although almost the feeblest of all the
entities of nature, simply by virtue of their tiny masses, they
seem to possess a subtle clue to some of the deepest laws of
nature pertaining to the unification-scale as well as the nature
of the unification-symmetry.  In this sense, the neutrinos provide
us with a rare window to view physics at truly short distances. As
we will see, these turn out to be as short as about $10^{-30}$~cm.
Furthermore, it appears most likely that the origin of their tiny
masses may be at the root of the origin of matter-antimatter
asymmetry in the early universe.  In short, the neutrinos may well
be crucial to our own origin!

The main purpose of my talk here today will be to present a unified picture,
in accord with observations, of
\begin{itemize}
\item Fermion masses and mixings
\item Neutrino oscillations
\item CP and flavor violations
\item Baryogenesis via leptogenesis, and
\item Proton decay
\end{itemize}
The goal will be to exhibit
the intimate links that exist between these different phenomena.
Each of these features, or a combination of some of them
(though not all in conjunction with each other),
have been considered widely in the literature.  My main
theme will be to exhibit how the first four, on which we have much empirical
data, hang \emph{together} within a single predictive framework based on the
gauge symmetry
\(G(224) = SU(2)_L \times SU(2)_R \times SU(4)^c\) \cite{JCPandAS} or
$SO(10)$ \cite{SO(10)}, leaving proton decay and supersymmetry as the two
missing pieces of this picture.  The crucial ingredients of the picture turn
out to be:
\begin{enumerate}
\item \textbf{Existence of the right-handed neutrino ($\nu_R$)} ---
that has been a compelling prediction of
the symmetry $SU(4)^c$ as well as of $SU(2)_L \times SU(2)_R$
from the 1970s \cite{JCPandAS}.
It is now needed to implement the seesaw mechanism \cite{seesaw} as well as
baryogenesis via leptogenesis \cite{FYKRS}.
\item \textbf{The observed meeting of the three gauge couplings} at a scale
\(M_U \approx 2 \times 10^{16} \mbox{\ GeV}\) \cite{susyunif}.
This observation on the one hand provides a strong evidence in
favor of the ideas of
both grand unification \cite{JCPandAS,GeorgiGlashow,GQW}
and supersymmetry \cite{susyWessZumino}; on the
other hand it sets the scale for the superheavy Majorana masses of the RH
neutrinos, which figure prominently in the seesaw formula for the masses of
the LH neutrinos.
\item \textbf{The gauge symmetry $SU(4)$-color}, which introduces three
\emph{characteristic features} ---of direct relevance to neutrino physics:
(a) the RH neutrinos (mentioned above); (b) B-L as a local
symmetry, which protects the RH neutrinos from acquiring a Planck or
string-scale mass; and (c) two simple mass-relations for the third family:
\begin{subequations}
\label{eq:1}
\begin{eqnarray}
m_{b}(M_{U}) & \approx & m_{\tau} \\
m(\nu_{\mathrm{Dirac}}^{\tau}) & \approx & m_{\mathrm{top}}(M_{U})
\end{eqnarray}
\end{subequations}
The first is successful empirically;
the second is crucial to the success of the seesaw formula
for $m(\nu_{L}^{\tau})$ which depends quadratically on
$m(\nu_{\mathrm{Dirac}}^{\tau})$ (see Sec.~4).  And,
\item \textbf{The seesaw mechanism} which combines the Dirac and the superheavy
Majorana masses of the neutrinos obtained as above to yield naturally light LH
neutrinos and in particular the right magnitude for $m(\nu_{L}^{\tau})$.
\end{enumerate}

To set the background for a discussion along these lines
I will first recall in the next section
the salient features of certain unification ideas based on $SU(4)$-color,
which developed in the early 1970s, their inter-relationships as well as
their relevance in the present context.  In a subsequent section, I will
present a predictive framework based on previous works on fermion masses and
neutrino oscillations, and in the following sections I will discuss the topics
of (a) CP and flavor violations, (b) leptogenesis, and (c) proton decay, as
they arise within the same framework.  In the last section I will present
a summary and make some concluding remarks.

\section{Unification with $SU(4)$-color: Neutrino Masses and the
Seesaw Mechanism} Going back to the days of Pauli, neutrinos have
been rather special from the day he postulated their existence in
the 1930s.  They were distinct then and are distinct even now from
all other known elementary particles in that they are essentially
massless and almost interactionless.  Up until the 1990s, prior to
the discoveries of neutrino oscillations, many (perhaps even most
in the 1970s) in fact believed, given that the upper limits on
neutrino masses were already known to be so small
(\(m(\nu_{e})/m_{e} \lsim 10^{-6}\) and, after the ``discovery''
of $\nu_{\tau}$, \(m(\nu_{\tau})/m_{\mathrm{top}} < 10^{-9}\)),
that the neutrinos will turn out to be exactly
massless.\footnote{The extent to which this belief was ingrained
among many (even in the 1990's) may be assessed by an interesting
remark by C. N. Yang at the recent Stony Brook Conference on
neutrinos \cite{Yang}.} This is in fact what the two component
theory of the neutrino \cite{lee} or the standard electroweak
model of particle physics \cite{GWS}, possessing only left-handed
neutrinos ($\nu_L$'s), would naturally suggest.\footnote{This is
barring, of course, possible contributions \cite{SWien}to the
Majorana mass of $\nu_L$ from lepton-number violating quantum
gravity effects \(\sim (v^{2}_{\mathrm{EW}}/M_{Pl}) \sim (250
\mbox{\ GeV})^2/10^{19} \mbox{\ GeV} \sim 10^{-5} \mbox{\ eV}\),
which are tiny, compared to the presently observed mass scales of
atmospheric and even solar neutrino oscillations. As expressed
elsewhere \cite{NewPati}, this smallness of (possible) quantum
gravity effects prompts one to regard the atmospheric and solar
neutrino oscillations as clear signals for physics beyond  the
standard model.} If neutrinos indeed existed only in the LH form
without a RH counterpart (a feature that would suggest itself if
neutrinos were truly massless), that would have implied that
nature is manifestly and \emph{intrinsically} left-right
asymmetric, parity violating. Many in the 1970s believed that that
may indeed be the case.  In fact the minimal grand unification
symmetry $SU(5)$ \cite{GeorgiGlashow} is built on such a belief.

There were, however, theoretical ideas of quark-lepton unification based on
the symmetry $SU(4)$-color and the concomitant idea of left-right
symmetry based on the commuting gauge symmetry
\(SU(2)_L \times SU(2)_R\), proposed in the early 1970s \cite{JCPandAS},
purely on aesthetic grounds, which professed that nature is intrinsically
quark-lepton and simultaneously left-right symmetric --- that is
parity-conserving.
Within this picture, the symmetries $SU(4)$-color \emph{and}
$SU(2)_R$ are assumed to be broken spontaneously at high energies
\cite{RNMandJCP},
such that $SU(3)$-color and $SU(2)_L$ remain unbroken; this stage of symmetry
breaking marks the onset of quark-lepton distinction and parity violation,
as observed at low energies.

Now the minimal symmetry containing $SU(4)$-color as well as the SM symmetry,
and simultaneously providing a compelling reason for the quantization of
electric charge is given by the symmetry group \cite{JCPandAS}:
\begin{equation}
G(224) \equiv SU(2)_L \times SU(2)_R \times SU(4)^c.
\end{equation}
Either one of the symmetries $SU(4)$-color or $SU(2)_R$ implies, however,
that \emph{there must exist the right-handed counterpart ($\nu_R$) of
the left-handed neutrino ($\nu_L$).}
This is because the RH neutrino ($\nu_R$) is the fourth color partner of
the RH up quark; and it is also the $SU(2)_R$
doublet partner of the RH electron.
Thus the symmetry $G(224)$
necessarily had to postulate the existence of an \emph{unobserved new
member in each family} --- the right-handed neutrino.  \emph{This requires
that there be sixteen two-component fermions in each family as opposed to
fifteen for the SM.}  Subject to left-right discrete symmetry
(L $\leftrightarrow$ R) which is natural to $G(224)$, all 16 members of
the electron family now became parts of a whole --- a single left-right
self-conjugate multiplet \(F = \{F_L \oplus F_R\}\), where
\begin{equation}
F_{L,R}^{e} = \left[ \begin{array}{cccc}
u_r & u_y & u_b & \nu_e \\
d_r & d_y & d_b & e^{-}
\end{array} \right]_{L,R} .
\end{equation}
The multiplets $F_{L}^{e}$ and $F_{R}^{e}$ are left-right conjugates of
each other transforming respectively as (2,1,4) and (1,2,4) of
$G(224)$; likewise for the muon and the tau families.  The symmetry
$SU(2)_{L,R}$ treat each column of $F_{L,R}^{e}$ as a doublet; while
the symmetry $SU(4)$-color unifies quarks and leptons by treating each
row $F_{L}^{e}$ \emph{and} $F_{R}^{e}$ as a quartet; \emph{thus lepton
number is treated as the fourth color.}  As mentioned above, because of the
parallelism between $SU(2)_L$ and $SU(2)_R$,
and because $SU(4)$-color is vectorial, the symmetry $G(224)$ naturally
permits the notion that the fundamental laws of nature possess a
left $\leftrightarrow$ right discrete symmetry (i.e.\ parity invariance)
that interchanges \(F_{L}^{e} \leftrightarrow F_{R}^{e}\) and
\(W_{L} \leftrightarrow W_{R}\); the observed parity violation is then
interpreted as being a low-energy phenomenon arising entirely through a
spontaneous breaking of the L $\leftrightarrow$ R discrete symmetry
\cite{RNMandJCP}.
I will return in just a moment to the relevance of having the RH neutrinos
for an understanding of the neutrino masses.  First, it is worth noting a
few additional features of the symmetry $G(224)$ and its relationship to still
higher symmetries.

The symmetry $G(224)$ introduces an elegant charge formula:
\(Q_{em} = I_{3L} + I_{3R} + (B-L)/2\), that applies to all
forms of matter (including quarks and leptons of all six flavors, Higgs and
gauge bosons).  Note that the quantum numbers of all members of a family,
including the weak hypercharge \(Y_W = I_{3R} + (B-L)/2\), are now completely
determined by the symmetry group $G(224)$ and the tranformation-property of
$(F_L \oplus F_R)$. This is in contrast to the case of the SM for which the
$15$ members of a family belong to five disconnected multiplets, with unrelated
quantum numbers. Quite clearly the charges $I_{3L}$, $I_{3R}$, and
$B-L$ being generators of $SU(2)_L$, $SU(2)_R$, and $SU(4)^c$ respectively
are quantized; so also then is the electric charge $Q_{em}$.

At this point, an intimate link between $SU(4)^c$ and
\(SU(2)_L \times SU(2)_R\) is worth noting.  Assuming that $SU(4)^c$
is gauged and demanding an explanation
of the quantization of electric charge as above
leaves one with no other choice but to gauge minimally
\(SU(2)_L \times SU(2)_R\) (rather than \(SU(2)_L \times U(1)_{I_{3R}}\)).
Likewise, assuming \(SU(2)_L \times SU(2)_R\) and again demanding a compelling
reason for the quantization of electric charge dictates that one must
minimally gauge $SU(4)^c$ (rather than
\(SU(3)^c \times U(1)_{B-L}\)).  The resulting minimal gauge symmetry is
then
\(G(224) = SU(2)_L \times SU(2)_R \times SU(4)^c\)
that simultaneously achieves quantization of electric charge, quark-lepton
unification and left-right symmetry \cite{JCPandAS}.  \emph{In short, the
concepts
of $SU(4)$-color and left-right gauge symmetry (symbolized by
\(SU(2)_L \times SU(2)_R\))
become inseparable from each other, if one demands that there be an
underlying reason for the quantization of electric charge.}  Assuming one
automatically implies the other.

In brief, the symmetry $G(224)$ brings some attractive features to particle
physics.  These include:
\begin{enumerate}
\item Unification of all 16 members of a family within \emph{one} left-right
self-conjugate multiplet;
\item Quantization of electric charge;
\item Quark-lepton unification through $SU(4)$-color and the consequent
mass relations given in Eq.~\ref{eq:1};
\item Conservation of parity at a fundamental level \cite{RNMandJCP};
\item Existence of the right-handed neutrinos ($\nu_R$'s) as a compelling
feature;
\item $B-L$ as a local symmetry; and
\item Just the right set-up (as was realized in the late 70's and 80's) for
implementing both the seesaw mechanism and leptogenesis.
\end{enumerate}
As I will discuss, the three features (3), (5), and (6) --- that
distinguish symmetries possessing $SU(4)$-color from alternative symmetries
like $SU(5)$ --- are now needed to provide the set-up mentioned in 7 and
thereby gain an understanding of the neutrino masses and the baryon-excess
of the universe.

Now one can retain all the advantages (1)--(7) of the symmetry group
$G(224)$ and in addition achieve gauge coupling unification,
if one extends the symmetry $G(224)$
(which is isomorphic
to \(SO(4) \times SO(6)\)) minimally into the
simple group $SO(10)$ \cite{SO(10)}.
As a historical note, it is worth noting, however, that \emph{all the
attractive features of $SO(10)$}, which distinguish it from $SU(5)$,
in particular the compelling need for the RH neutrino, the L-R discrete
symmetry,\footnote{In the context of $SO(10)$ the L-R discrete symmetry
associated with $G(224)$ is replaced by an equivalent D-parity symmetry.
See Ref.~\cite{Chang}.}
$B-L$ and $SU(4)$-color symmetry, which are now relevant to an
understanding of the neutrino masses and baryon asymmetry,
\emph{were introduced entirely through the symmetry $G(224)$}
\cite{JCPandAS}, long before the $SO(10)$ papers appeared.
This is because these
features arise already at the level of the symmetry $G(224)$.  The symmetry
$SO(10)$ of course fully preserves these features because it
contains $G(224)$ as a subgroup.  It is furthermore remarkable that
$SO(10)$ preserves even the left-right
conjugate
16-plet multiplet structure of $G(224)$ by using the set
\(F = (F_L \oplus (F_R)^c)\) to represent the members of a family.
The 16-plet now constitutes the sixteen-dimensional spinorial
representation of $SO(10)$.
Thus
$SO(10)$ does not need to add any new matter-fermions beyond those of
$G(224)$.  By contrast, if one extends $G(224)$ to $E_6$ \cite{E6},
the advantages (1)--(7) are retained but in this case one must extend the
family structure from a 16 to a 27-plet by postulating additional
fermions.

In contrast to the extension of $G(224)$ to $SO(10)$ or $E_6$, if one
wished to extend only the SM symmetry $G(213)$ to a simple group, the
minimal such extension would be $SU(5)$ \cite{GeorgiGlashow}.  In the 1970s,
long before
the discovery of neutrino oscillations, the symmetry $SU(5)$, being the
smallest simple group possessing the SM symmetry, had the virtue of
demonstrating the ideas of grand unification simply.  It, however, does not
contain $G(224)$ as a subgroup.  As such, except for quantization of
electric charge (feature (2)), $SU(5)$ does not possess any of the other
features (i.e.\ features (1), (3), (4), (5), (6) and (7)) of $G(224)$
listed above.
In particular, it does not contain (a) the RH neutrino as a compelling
feature, (b) $B-L$ as a local symmetry, and (c) the second mass-relation
of Eq.~\ref{eq:1}, based on $SU(4)$-color.\footnote{Like $SO(10)$,
$SU(5)$ does possess, however, fermion to anti-fermion gauge transformations
which are absent in $G(224)$.}
As discussed below, all three
of these features play crucial roles in providing an understanding of
neutrino masses and in implementing baryogenesis via leptogenesis.
Furthermore $SU(5)$
splits members of a family (not including $\nu_R$ or $(\nu_R)^c$) into
two multiplets: \(\bar{5}+10\).  And it violates parity, like the SM,
manifestly.

In short, the symmetries $SO(10)$ and $E_6$ possess all the advantages
(1)--(7) listed above because they contain $G(224)$ as a subgroup, while
$SU(5)$ does not possess them (barring feature (2)) because it does not
contain $G(224)$ as a subgroup.

Having discussed some of the main ideas on higher unification,
which developed in the early 1970s, I now return to a discussion
of the issue of the neutrino masses, that arises in this context
and the need for the seesaw mechanism.  As we saw, symmetries
based on either $SU(4)$-color or left-right symmetry implies that
the LH neutrinos ($\nu_L$'s) must have their RH counterparts (the
$\nu_R$'s).  That in turn implies, however, that the neutrino
should acquire at least a Dirac mass, in short \emph{it must be
massive (not massless).}  The dilemma that faced such a theory in
the early 1970s, with the RH neutrino being linked to the RH up
quark through $SU(4)$-color, despite its aesthetic merits, is
however this: What makes the neutrino so extra-ordinarily light
($\lsim 1\mbox{\ eV}$) compared to the other fermions, including
even the electron?\footnote{Note that neither $SU(5)$ nor the
standard model faced this dilemma from the start (especially when
people generally believed in the masslessness of the neutrinos)
because in either case there is no need for the RH neutrinos. And
with only LH neutrinos, the neutrinos can remain exactly massless
as long as $B-L$ is conserved. (Possible violations of lepton
number leading to \(\Delta L = \pm 2\) and \(\Delta B = 0\)
operators arising through quantum-gravity effects can make only
tiny contributions ($\sim 10^{-5}\mbox{\ eV}$) to the Majorana
masses of the neutrinos, as mentioned in Footnote~2).  For this
reason, believing in massless neutrinos, many had expressed in the
1970s their preference for $SU(5)$ over $G(224)$ or $SO(10)$.  In
their opinion, the RH neutrino was an unnecessary and uneconomical
luxury.  One's only defense at that time was reliance on
aesthetics; the ideas of $SU(4)$-color and left-right symmetry
appeared (to me) to be much prettier.  As is well known now and as
discussed below, with the invention of the seesaw mechanism and
the discovery of neutrino oscillations, the situation has changed
dramatically. \emph{The RH neutrino, introduced in the early
1970s, is no longer a luxury but a necessity.}} Although the
resolution of this dilemma was staring at one's face, given that
the RH neutrinos are singlets of the SM and that $B-L$ is
necessarily violated\footnote{Once $B-L$ is gauged and thus
coupled to a massless gauge boson, such a gauge boson must acquire
a mass through SSB so as to avoid conflict with the E\"otvos-type
experiments.  In this case, $B-L$ must be violated spontaneously
\cite{JCPandAS}.} in a theory as proposed in \cite{JCPandAS}, it
waited for six years until 1979 when the seesaw mechanism was
discovered \cite{seesaw}, which fully resolved the dilemma.

The idea of the seesaw mechanism is simply this.  In a theory with spontaneous
breaking of $B-L$ and $I_{3R}$ at a high scale ($M$), already inherent in
\cite{JCPandAS}, the RH neutrinos can and generically will acquire a superheavy
Majorana mass ($M(\nu_R) \sim M$) that violates lepton number and $B-L$
by two units.  Combining this with the Dirac mass of the neutrino
($m(\nu_{\mathrm{Dirac}})$),
which arises through electroweak symmetry breaking,
one would then obtain a mass for the LH neutrino given by
\begin{equation}
m(\nu_L) \approx m(\nu_{\mathrm{Dirac}})^2/M(\nu_R)
\label{eq:4}
\end{equation}
which would be naturally super-light because $M(\nu_R)$ is naturally
superheavy.  This then provided a \emph{simple but compelling reason} for the
lightness of the known neutrinos.  In turn it took away the major burden that
faced the ideas of $SU(4)$-color and left-right symmetry from the beginning.
In this sense, the seesaw mechanism was indeed the \emph{missing piece} that
was needed to be found for consistency of the ideas of $SU(4)$-color and
left-right symmetry.

In turn, of course, the seesaw mechanism needs the ideas of $SU(4)$-color
and SUSY grand unification so that it may be quantitatively useful.  Because
these two ideas not only provide (a) the RH neutrino as a compelling
feature (crucial to seesaw), but also provide respectively (b) the Dirac mass
for the tau neutrino (cf.\ Eq.~\ref{eq:1}), and (c) the superheavy Majorana
mass of the $\nu_{R}^{\tau}$ (see Sec.~4).  Both these masses enter crucially
into the seesaw formula and end up giving the \emph{right mass-scale} for
the atmospheric neutrino oscillation as observed.
To be specific, Eq.~(\ref{eq:1}) based on $SU(4)$-color yields
\(m(\nu_{\mathrm{Dirac}}^{\tau}) \approx m_{\mathrm{top}}(M_U)
\approx 120 \mathrm{\ GeV}\),
and the SUSY unification scale, together with the protection provided by
$B-L$ that forbids Planck-scale contributions to the Majorana mass of
$\nu_{R}^{\tau}$, naturally yields
\(M(\nu_{R}^{\tau}) \sim 10^{15}\mathrm{\ GeV}(1/2\mbox{--}2)\)
[cf. Sec.~4].  The seesaw formula (\ref{eq:4}) generalized to include
2-3 family mixing then yields
\(m(\nu_{L}^{3}) \approx
(2.9)(120\mathrm{\ GeV})^{2}/10^{15}\mathrm{\ GeV}(1/2\mbox{--}2)
\approx (1/24\mathrm{\ eV})(1/2\mbox{--}2)\), where the factor 2.9
comes from 2-3 mixing (see Sec.~4).  This is just the right
magnitude to go with the mass scale observed at SuperK \cite{sk}!

\emph{Without an underlying reason as above for at least the
approximate values of these two vastly differing mass-scales ---
$m(\nu_{\mathrm{Dirac}}^{\tau})$ and $M(\nu_{R}^{\tau})$ --- the
seesaw mechanism by itself would have no clue, quantitatively, to
the mass of the LH neutrino.}  In fact it would yield a rather
arbitrary value for $m(\nu_{L}^{\tau})$, which could vary quite
easily by more than 10 orders of magnitude either way around the
observed mass scale.\footnote{To see this, consider for simplicity
just the third family.  Without $SU(4)$-color, even if a RH
two-component fermion $N$ (the analogue of $\nu_R$) is introduced
by hand as a \emph{singlet} of the gauge symmetry of the SM or
$SU(5)$, \emph{such an $N$ by no means should be regarded as a
member of the third family, because it is not linked by a gauge
transformation to the other fermions in the third family}.  Thus
its Dirac mass term given by \(m(\nu_{\mathrm{Dirac}}^{\tau})[
\bar{\nu}_{L}^{\tau} N + h.c. ]\) is completely arbitrary, except
for being bounded from above by the electroweak scale $\sim 200
\mbox{\ GeV}$. In fact a priori (within the SM or $SU(5)$) it can
well vary from say 1~GeV (or even 1~MeV) to 100~GeV. Using
Eq.~(\ref{eq:4}), this would give a variation in $m_(\nu_L)$ by at
least four orders of magnitude if the Majorana mass $M(N)$ of $N$
is held fixed.  Furthermore, $N$ being a singlet of the SM as well
as of $SU(5)$, the Majorana mass $M(N)$, unprotected by $B-L$,
could well be as high as the Planck or the string scale
(\(10^{18}\mbox{--}10^{17}\mbox{\ GeV}\)), and as low as say
1~TeV; this would introduce a further arbitrariness (by some
fourteen orders of magnitude) in $m(\nu_L)$. Such arbitrariness
both in the Dirac and in the Majorana masses, is drastically
reduced, however, once $\nu_R$ is related to the other fermions in
the family by an $SU(4)$-color gauge transformation (see
Eq.~(\ref{eq:1}) and Sec.~4).}

In short, the seesaw mechanism needs the ideas of SUSY unification and
$SU(4)$-color, and of course vice-versa; \emph{together} they provide
an understanding
of neutrino masses as observed.  Schematically, one thus finds:
\begin{equation}
\begin{array}{rcl}
\fbox{\(
\begin{array}{c}
\mbox{SUSY UNIFICATION} \\ \mbox{WITH $SU(4)$-COLOR}
\end{array}\)} & \oplus & \fbox{SEESAW} \\
 & \Downarrow & \\
m(\nu_{L}^{3}) & \sim & 1/10 \mbox{\ eV}.
\end{array}
\label{eq:scheme}
\end{equation}

I will return to a more quantitative discussion of the mass scale and the
angle associated with the atmospheric neutrino oscillations in Sec.~4.
In the next section I first briefly discuss (assuming a string-theoretic
origin of the effective symmetry in four dimensions) the issue of the
four-dimensional symmetry being either $G(224)$ or $SO(10)$ near the
string scale.

\section{The Question of a Possible Preference For
the Effective Symmetry in 4D Being $G(224)$ or $SO(10)$}
We have argued in the previous section that one needs an effective symmetry
in 4D like $G(224)$ or $SO(10)$ containing $SU(4)$-color to understand
neutrino masses.  Such a need will be further strengthened by our discussions
of fermion masses, neutrino oscillations, and leptogenesis in Secs.~4 and 6.
The advantages of these two symmetries --- $G(224)$ and $SO(10)$ --- as regards
these three issues turn out to be rather identical.
Here I briefly present some characteristic differences between the two
symmetries --- $G(224)$ versus $SO(10)$ --- and discuss the question of
whether one may have a good reason to choose between them, viewing each
of these as an effective symmetry in 4D, that emerges from an underlying theory
like the string/M theory in higher dimensions \cite{stringth}. The answer
depends in part on an understanding of the observed gauge coupling unification
on the one hand and resolving the problem of doublet-triplet splitting
for SUSY GUT-theories on the other hand.

It has been known for some time that when the three gauge couplings are
extrapolated from their values measured at LEP to higher energies,
in the context of weak-scale supersymmetry \cite{susyunif}\footnote{The case of
weak-scale supersymmetry is of course motivated independently by the desire to
avoid unnatural fine tuning in the Higgs mass.}, they meet, to a very good
approximation, at a scale given by:
\begin{equation}
M_U \approx 2\times 10^{16} \mbox{\ GeV}
\end{equation}
This dramatic meeting of the three gauge couplings provides a strong support
for the ideas of both grand unification and supersymmetry.

The most straightforward interpretation of such a meeting of the gauge
couplings is that a supersymmetric grand unification symmetry (often called
GUT symmetry), like $SU(5)$ or $SO(10)$, is operative above the scale
$M_U$ and that it breaks spontaneously into the SM symmetry $G(213)$ at
around $M_U$, while supersymmetry breaking occurs at some high scale and
induces soft masses of order one TeV.

Even if supersymmetric grand unification is a good effective theory below
a certain scale $M$ lying above $M_U$ (\(M_U \lsim M\)), it seems
imperative, however, that it should have its origin within an underlying
theory like the string or $M$-theory, which is needed
to provide a good quantum theory of gravity and also to unify all the
forces of nature including gravity.

In the context of string or $M$-theory defined in $D=10$ or $11$, an
alternative interpretation of the meeting of the three gauge couplings is,
however, possible.  This is because even if the effective symmetry in 4D
emerging from the string theory is non-simple like $G(224)$, string theory
can still ensure familiar gauge coupling unification at the string-scale
$M_{\mathrm{st}}$ \cite{Ginspang,DienesJCP}.
With this in mind one can consider two
alternative possibilities both of which would account for coupling unification
and would also be equally suitable for understanding neutrino masses and
leptogenesis.

First, if the effective symmetry in 4D emerging
from the string/$M$-theory is a simple group like $SO(10)$, which breaks
into the SM symmetry at $M_U$ by the Higgs mechanism, the observed gauge
coupling unification can of course be understood simply in this case,
even if $M_{\mathrm{st}}$ is (say) an order of magnitude higher than $M_U$.
This is because $SO(10)$ will preserve coupling unification from
$M_{\mathrm{st}}$ down to $M_U$.\footnote{With the GUT symmetry $SO(10)$
being intact above $M_U$, one should still ensure that
the $SO(10)$ gauge coupling
does not grow too rapidly to become non-perturbative
(\(\alpha_{\mathrm{GUT}} \gsim 1\))
below $M_{\mathrm{st}}$.  (This would
in fact suggest avoiding large Higgs multiplets like 126 and
$\overline{126}$)}

Second, even if the effective symmetry
in 4D emerging from string theory is non-simple like $G(224)$ (as in
\cite{antoniadisLeon}), as long as the
string-scale is not far above the GUT-scale
(suppose \(M_{\mathrm{st}} \approx (2\mbox{--}3)M_U\),
say\footnote{The case of the string scale being rather close to the GUT
scale as above can arise quite plausibly by utilizing the ideas of string
duality (following E.~Witten \cite{WittenDual}) which can lower the string
scale below its perturbative value of
\(\approx 4 \times 10^{17}\mathrm{\ GeV}\) \cite{Ginspang},
and/or those of
semi-perturbative unification (K.\ S.\ Babu and J.\ C.\ Pati
\cite{BabuJi})
which raises $M_{\mathrm{GUT}}$ above the conventional MSSM value of
\(2 \times 10^{16}\mathrm{\ GeV}\).}),
the couplings of $G(224)$ unified at the string scale will remain essentially
so at the GUT scale ($M_U$) so as to match the observed coupling unification.
Despite the short gap between $M_{\mathrm{st}}$ and $M_U$ in this case,
one would still have the benefits of $SU(4)$-color to understand neutrino
masses (as alluded to in Sec.~2) and baryogenesis via leptogenesis (to be
discussed in Sec.~6).  In short,
observed coupling unification can be attributed to either a simple group
like $SO(10)$ or a string-derived non-simple group like $G(224)$ (with
\(M_{\mathrm{st}} \approx (2\mbox{--}3)M_U\)) being effective in 4D above
the GUT scale $M_U$.

There is, however, a characteristic difference between a GUT (like $SO(10)$)
versus a non-GUT (like $G(224)$) string solution in 4D as follows.  A SUSY
4D GUT solution possessing symmetries like $SO(10)$ would need the color
triplets in the $10_H$ of $SO(10)$ (see Sec.~4) to become superheavy, while
doublets remain light, by the so-called doublet-triplet splitting
mechanism, so as to avoid rapid proton decay.  While such a mechanism can
be constructed for a 4D theory \cite{DimWil}, it requires a rather special
choice of Higgs multiplets and of their couplings, and it remains to be
seen whether such a choice can in fact emerge for a string-derived $SO(10)$
solution in 4D.

Non-GUT string solutions (based on symmetries like
$G(224)$ or $G(2113)$) have a distinct advantage in this regard over a
SUSY GUT solution in that the dangerous color triplets that induce rapid
proton decay are often naturally projected out for these solutions
\cite{Candelas,antoniadisLeon,Faraggi}.\footnote{One must still ensure
in the context of a realistic $G(224)$ (or $SO(10)$) solution, capable
generating CKM mixings, that only one pair of Higgs doublets ($Hu$ and
$Hd$) remain light.  A possible mechanism for realizing such a solution
is noted in \cite{BPW1}.}
Furthermore the non-GUT solutions invariably yield
desired ``flavor'' symmetries, which help resolve certain
naturalness problems of supersymmetry such as those pertaining to the issues
of squark degeneracy \cite{FaraggiJCP},
CP violation \cite{BabuJCP}, and quantum gravity
induced rapid proton decay \cite{JCPProton}.
I should mention that promising string theory solutions yielding the
$G(224)$-symmetry in 4D have been obtained (using different approaches) by
a number of authors \cite{StringG(224)}.  And, recently there have also been
several attempts based on compactifications of five and six-dimensional
GUT-theories which yield the $G(224)$-symmetry in 4D with some very
desirable features \cite{5DG(224)}.

Weighing the advantages and possible disadvantages of both, it seems hard at
present to make a priori a clear choice between a GUT $SO(10)$ solution or a
non-GUT $G(224)$ solution emerging from string theory in 4D.  As expressed
elsewhere \cite{JCPEriceReview},
it therefore seems prudent to keep both options open
and pursue their phenomenological consequences.  As mentioned in Sec.~2 and
discussed further in the following sections, the advantages of both are
essentially the same as regards gaining an understanding of fermion masses,
neutrino oscillations, and baryongenesis via leptogenesis.
In Sec.~8, distinctions
between the two cases as regards proton decay will be noted.

\section{Fermion Masses and Neutrino Oscillations Within a
$G(224)/SO(10)$-Framework}
Following Ref.~\cite{BPW1}, I now present a simple and predictive
pattern for fermion mass-matrices based on $SO(10)$ or the
$G(224)$-symmetry.\footnote{I will present the Higgs system for
$SO(10)$.  The discussion would remain essentially unaltered if
one uses the corresponding $G(224)$-submultiplets instead.} One
can obtain such a mass mass-matrix for the fermions by utilizing
only the minimal Higgs system that is needed to break the
gauge symmetry $SO(10)$ to \(SU(3)^{c} \times U(1)_{em}\).  It
consists of the set:
\begin{equation}
H_{\mathrm{minimal}} =
\left\{ \mathbf{45_{H}}, \mathbf{16_{H}},
\overline{\mathbf{16}}_{\mathbf{H}}, \mathbf{10_{H}} \right\}
\label{Hmin}
\end{equation}
Of these, the VEV of \(\left\langle \mathbf{45_{H}} \right\rangle
\sim M_{X}\) breaks $SO(10)$ in the B-L direction to \(G(2213) =
SU(2)_{L} \times SU(2)_{R} \times U(1)_{B-L} \times SU(3)^{c}\),
and those of \(\left\langle \mathbf{16_{H}} \right\rangle =
\left\langle \overline{\mathbf{16}}_{\mathbf{H}} \right\rangle\)
along \(\left\langle \tilde{\bar{\nu}}_{RH} \right\rangle\) and
\(\left\langle \tilde{\nu}_{RH} \right\rangle\)\ break $G(2213)$
into the SM symmetry $G(213)$ at the unification-scale $M_{X}$.
Now $G(213)$ breaks at the electroweak scale by the VEV of
\(\left\langle \mathbf{10_{H}} \right\rangle\) to \(SU(3)^{c}
\times U(1)_{em}\).\footnote{Large dimensional tensorial
multiplets of $SO(10)$ like $\mathbf{126}_{H}$,
$\overline{\mathbf{126}}_{H}$, $\mathbf{120}_{H}$, and
$\mathbf{54}_{H}$ are not used for the purpose in part because they tend
to give too large threshold corrections to $\alpha_{3}(m_{Z})$
(typically exceeding 20\%), which would render observed coupling
unification fortuitous [see e.g. discussions in Appendix D of
Ref.~\cite{BPW1}].  Furthermore, the multiplets like $126_{H}$ and
$120_{H}$ do not seem to arise at least in weakly interacting heterotic
string solutions \cite{DienesMarch}.}

The $3\times 3$ Dirac mass matrices for the four sectors $(u,d,l,\nu)$
proposed in Ref. \cite{BPW1} were motivated in part by the notion that flavor
symmetries \cite{Flavorsymm} are responsible
for the hierarchy among the elements
of these matrices (i.e., for
\(\mbox{``}33\mbox{''} \gg \mbox{``}23\mbox{''}
\gg \mbox{``}22\mbox{''} \gg \mbox{``}12\mbox{''}
\gg \mbox{``}11\mbox{''}\), etc.),
and in part by the group theory of $SO(10)/G(224)$, relevant to a minimal Higgs
system (see below). Up to minor
variants,\footnote{The zeros in ``11'', ``13'', and ``31''
elements signify that they
are relatively small quantities (specified below). While the ``22' elements
were set to zero in Ref. \cite{BPW1}, because they are meant to be
\(< \mbox{``}23\mbox{''} \mbox{``}32\mbox{''}/\mbox{''}33\mbox{''}
\sim 10^{-2}\)
(see below), and thus unimportant for purposes
of Ref. \cite{BPW1}, they are retained here, because such small
$\zeta_{22}^u$ and $\zeta_{22}^d$ [$\sim (1/3)\times 10^{-2}$ (say)] can
still be important for CP violation and thus leptogenesis.}
they are as follows:\footnote{A somewhat analogous pattern, also based
on $SO(10)$, has
been proposed by C. Albright and S. Barr [AB] \cite{AlbrightBarr}. One
major difference between the work of AB and that of BPW
\cite{BPW1} is that the former introduces the so-called
``lop-sided'' pattern in which some of the ``23'' elements are even
greater than the ``33'' element; in BPW on the otherhand, the
pattern is consistently hierarchical with individual ``23'' elements
(like $\eta$, $\epsilon$ and $\sigma$) being much smaller in
magnitude than the ``33'' element of 1. For a comparative study of some
of the $SO(10)$-models for fermion masses and neutrino oscillations and
the corresponding references, see C.H.~Albright, talk presented at the
Stony Brook conf.\ (Oct.\ 2002), Ed.\ by R.\ Shrock, publ.\ by
World Scientific (page 201).}
\begin{subequations}
\label{eq:mat}
\begin{eqnarray}
M_u & = &\left[
\begin{array}{ccc}
0&\epsilon'&0\\-\epsilon'&\zeta_{22}^u&\sigma+\epsilon\\0&\sigma-\epsilon&1
\end{array}\right] \mathcal{M}_u^0 \\
M_d & = & \left[
\begin{array}{ccc}
0&\eta'+\epsilon'&0\\ \eta'-\epsilon'&\zeta_{22}^d&\eta+\epsilon\\0&
\eta-\epsilon&1
\end{array}\right] \mathcal{M}_d^0 \\
M_\nu^D & = &\left[
\begin{array}{ccc}
0&-3\epsilon'&0\\3\epsilon'&\zeta_{22}^u&\sigma-3\epsilon\\
0&\sigma+3\epsilon&1\end{array}\right] \mathcal{M}_u^0 \\
M_l & = & \left[
\begin{array}{ccc}
0&\eta'-3\epsilon'&0\\ \eta'+3\epsilon'&\zeta_{22}^d&\eta-3\epsilon\\0&
\eta+3\epsilon&1
\end{array}\right]\mathcal{M}_d^0 \nonumber
\end{eqnarray}
\end{subequations}
These matrices are defined in the gauge basis and are
multiplied by $\bar\Psi_L$ on left and $\Psi_R$ on right.
For instance, the row and column indices of $M_u$ are given by $(\bar u_L,
\bar c_L, \bar t_L)$ and $(u_R, c_R, t_R)$ respectively.
Note the group-theoretic up-down and quark-lepton correlations: the same
$\sigma$ occurs in $M_u$ and $M_\nu^D$, and the same $\eta$ occurs in $M_d$
and $M_l$. It will become clear that the $\epsilon$ and $\epsilon'$ entries
are proportional to $B-L$ and are antisymmetric in the family space
(as shown above).
Thus, the same $\epsilon$ and $\epsilon'$ occur in both ($M_u$ and $M_d$) and
also in ($M_\nu^D$ and $M_l$), but $\epsilon\rightarrow -3\epsilon$ and
$\epsilon'\rightarrow -3\epsilon'$ as $q\rightarrow l$. Such correlations
result in enormous reduction of parameters and thus in increased predictivity.
Such a pattern for the mass-matrices can be obtained, using a minimal Higgs
system \(\mathbf{45}_H\), \(\mathbf{16}_H\),
\(\overline{\mathbf{16}}\), and \(\mathbf{10}_H\)
and a singlet $S$ of $SO(10)$, through effective couplings as
follows \cite{FN26}:
\begin{eqnarray}
\label{eq:Yuk}
\lefteqn{\mathcal{L}_{\mathrm{Yuk}} =} & & \nonumber \\
& & h_{33}\mathbf{16}_3 \mathbf{16}_3 \mathbf{10}_H \nonumber \\
 & & \mbox{} + h_{23} \mathbf{16}_2 \mathbf{16}_3 \mathbf{10}_H(S/M)
\nonumber \\
 & & \mbox{} + a_{23} \mathbf{16}_2 \mathbf{16}_3 \mathbf{10}_H
(\mathbf{45}_H/M')(S/M)^p \nonumber \\
 & & \mbox{} + g_{23}\mathbf{16}_2 \mathbf{16}_3 \mathbf{16}_H^d
(\mathbf{16}_H/M'')(S/M)^q \nonumber \\
 & & \mbox{} + h_{22} \mathbf{16}_2 \mathbf{16}_2 \mathbf{10}_H(S/M)^2 \\
 & & \mbox{} + g_{22} \mathbf{16}_2 \mathbf{16}_2
\mathbf{16}_H^d(\mathbf{16}_H/M'')(S/M)^{q+1} \nonumber \\
 & & \mbox{} + g_{12} \mathbf{16}_1 \mathbf{16}_2
\mathbf{16}_H^d(\mathbf{16}_H/M'')(S/M)^{q+2} \nonumber \\
 & & \mbox{} + a_{12}\mathbf{16}_1 \mathbf{16}_2
\mathbf{10}_H( \mathbf{45}_H/M')(S/M)^{p+2} \nonumber
\end{eqnarray}
Typically we expect $M'$, $M''$ and $M$ to be of order $M_{\mathrm{string}}$
\cite{FN6}. The VEV's of \(\langle\mathbf{45}_H\rangle\) (along $B-L$),
\(\langle \mathbf{16}_H\rangle = \langle\overline{\mathbf{16}}_H\rangle\)
(along standard model singlet sneutrino-like component) and of
the $SO(10)$-singlet $\langle S \rangle$
are of the GUT-scale, while those of $\mathbf{10}_H$ and of the down type
SU(2)$_L$-doublet component in $\mathbf{16}_H$ (denoted by $\mathbf{16}_H^d$)
are of the electroweak scale \cite{BPW1,FN7}. Depending upon whether
\(M'(M'')\sim M_{\mathrm{GUT}}\) or $M_{\mathrm{string}}$
(see \cite{FN8}), the
exponent $p(q)$ is either one or zero \cite{FN8}.

The entries 1 and $\sigma$ arise respectively from $h_{33}$ and $h_{23}$
couplings, while $\hat\eta\equiv\eta-\sigma$ and $\eta'$ arise respectively
from $g_{23}$ and $g_{12}$-couplings. The (B-L)-dependent antisymmetric
entries $\epsilon$ and $\epsilon'$ arise respectively from the $a_{23}$ and
$a_{12}$ couplings. [Effectively, with
\(\langle\mathbf{45}_H\rangle\propto B-L\),
the product $\mathbf{10}_H\times\mathbf{45}_H$ contributes as a $\mathbf{120}$,
whose coupling
is family-antisymmetric.] The small entry $\zeta_{22}^u$ arises from the
$h_{22}$-coupling, while $\zeta_{22}^d$ arises from the joint contributions of
$h_{22}$ and $g_{22}$-couplings. As discussed in \cite{BPW1}, using some of
the observed masses as inputs, one obtains
\(|\hat{\eta}| \sim |\sigma| \sim |\epsilon| \sim \mathcal{O}(1/10)\),
$|\eta'|\approx 4\times 10^{-3}$ and $|\epsilon'|\sim 2\times 10^{-4}$. The
success of the framework presented in Ref. \cite{BPW1}
(which set $\zeta_{22}^u=\zeta_{22}^d=0$) in describing fermion masses and
mixings remains essentially unaltered if
$|(\zeta_{22}^u,\zeta_{22}^d)|\leq (1/3)(10^{-2})$ (say).

Such a hierarchical form of the mass-matrices, with $h_{33}$-term being
dominant, is attributed in part to flavor gauge symmetry(ies) that
distinguishes between the three families, and in part to higher
dimensional operators involving for example $\langle\mathbf{45}_H\rangle/M'$
or $\langle\mathbf{16}_H\rangle/M''$, which are suppressed by
\( M_{\mathrm{GUT}}/M_{\mathrm{string}} \sim 1/10\), if $M'$ and/or
\(M''\sim M_{\mathrm{string}}\).
The basic presumption here is that effective dimensionless
couplings allowed by $SO(10)/G(224)$ and flavor symmetries are of order unity
[i.e., $(h_{ij},g_{ij},a_{ij})\approx 1/3$-3 (say)]. The need for appropriate
powers of $(S/M)$ with
\(\langle S\rangle/M \sim M_{\mathrm{GUT}}/M_{\mathrm{string}}
\sim (1/10\mbox{--}1/20)\)
in the different couplings leads to a hierarchical structure.
As an example, introduce just one U(1)-flavor symmetry,
together with a discrete symmetry $D$,
with one singlet $S$. The
hierarchical form of the Yukawa couplings exhibited in Eqs.
(\ref{eq:mat}) and  (\ref{eq:Yuk}) would
follow, for the case of $p=1$, $q=0$, if,
for example, the $U(1)$ flavor charges are assigned as follows:
\begin{eqnarray}
\begin{array}{cccc}
\mathbf{16}_3  & \mathbf{16}_2 & \mathbf{16}_1 & \mathbf{10}_H  \\
a & a+1 & a+2 & -2a
\end{array} \nonumber \\
\begin{array}{cccc}
\mathbf{16}_H & \overline{\mathbf{16}}_{H} & \mathbf{45}_H & \mathbf{S} \\
-a-1/2 & -a & 0 & -1
\end{array}
\label{eqn:charges}
\end{eqnarray}
The value of $a$ would get fixed by the presence of other
operators (see later).
All the fields are assumed to be even under the discrete symmetry $D$,
except for $\mathbf{16}_{H}$ and $\overline{\mathbf{16}}_{H}$ which are odd.
It is assumed that other fields are present that would make the U(1) symmetry
anomaly-free. With this assignment of charges, one
would expect
\(|\zeta_{22}^{u,d}|\sim (\langle S\rangle/M)^2\); one may thus take,
for example,
\(|\zeta_{22}^{u,d}|\sim (1/3)\times 10^{-2}\) without upsetting the success of
Ref.~\cite{BPW1}. In the same spirit, one would expect $|\zeta_{13}$,
\(\zeta_{31}| \sim (\langle S\rangle/M)^2\sim 10^{-2}\), and
\(|\zeta_{11}|\sim (\langle S\rangle/M)^4\sim 10^{-4}\) (say).
where $\zeta_{11}$, $\zeta_{13}$, and $\zeta_{31}$ denote the ``11,''
``13,'' and ``31,'' elements respectively.
In the interest of economy in parameters and thus greater predictivity,
we drop these elements ($\zeta_{11}$, $\zeta_{13}$, $\zeta_{31}$, and
even $\zeta_{22}$) as a first approximation, in this section, as in
Ref.\ \cite{BPW1}.  But these elements can in general be relevant in a more
refined analysis (e.g.\ $\zeta_{11}^{u,d}$, though small, can make small
contributions to $m_{u,d}$ of order few MeV without altering significantly
the mixing angles, and $\zeta_{22}$ can be relevant for considerations of
CP violation).

To discuss the neutrino sector one must specify the Majorana mass-matrix of
the RH neutrinos as well. These arise from the effective couplings of the
form
\cite{FN30}:
\begin{equation}
\label{eq:LMaj}
\mathcal{L}_{\mathrm{Maj}} = f_{ij} \mathbf{16}_i \mathbf{16}_j
\overline{\mathbf{16}}_H \overline{\mathbf{16}}_H/M
\end{equation}
where the $f_{ij}$'s include appropriate powers of $\langle S \rangle/M$, in
accord with flavor charge assignments of $\mathbf{16}_i$
(see Eq.~(\ref{eqn:charges})), and $M$ is expected to be of order string or
reduced Planck scale.
For the $f_{33}$-term to be leading ($\sim 1$), we must assign
the charge $-a$ to
$\overline{\mathbf{16}}_H$. This leads to a hierarchical form for the Majorana
mass-matrix
\cite{BPW1}:
\begin{eqnarray}
\label{eq:MajMM}
M_R^\nu=\left[
\begin{array}{ccc}
x & 0 & z \\
0 & 0 & y \\
z & y & 1
\end{array}
\right]M_R
\end{eqnarray}
Following the flavor-charge assignments given in Eq.~(\ref{eqn:charges}), we
expect \(|y|\sim \langle S/M\rangle\sim 1/10\),
\(|z| \sim (\langle S/M\rangle)^2 \sim (1/200)(1 \mbox{ to } 1/2\), say),
\(|x| \sim (\langle S/M\rangle)^4 \sim (10^{-4}\mbox{--}10^{-5})\) (say).
The ``22'' element (not shown) is \(\sim (\langle S/M\rangle)^2\) and its
magnitude is
taken to be $< |y^2/3|$, while the ``12'' element (not shown) is
\(\sim (\langle S/M\rangle)^3\).
In short, with the assumption that the ``33''-element is leading,
\emph{the hierarchical pattern of $M_{R}^{\nu}$ is identical to that of
the Dirac mass matrices (Eq.~(8)).}  We expect
\begin{equation} \label{MR}
M_R = \frac{f_{33} \langle \overline{\mathbf{16}}_H \rangle^2}{M}
\approx (10^{15}\mbox{ GeV})(1/2\mbox{--}2)
\end{equation}
where we have put \(\langle\overline{\mathbf{16}}_H \rangle
\approx 2 \times 10^{16}\mbox{\ GeV}\),
\(M \approx M_{\mathrm{string}}\approx 4\times 10^{17}\mbox{\ GeV}\)
\cite{Ginspang} and  $f_{33}\approx 1$.
These lead to an expected central value of $M_R$ of around
$10^{15}\mathrm{\ GeV}$.
Allowing for 2-3 family-mixing in the
Dirac and the Majorana sectors as in
Eqs. (7) and (10), the seesaw mechanism leads to \cite{BPW1}:
\begin{equation}
m(\nu_{3}) \approx B \frac{m(\nu_{\mathrm{Dirac}}^{\tau})^{2}}{M_{R}}
\label{seesaw}
\end{equation}
The quantity $B$ represents the effect of 2-3 family-mixing and is given
by \(B = (\sigma + 3\epsilon)(\sigma + 3\epsilon - 2y)/y^{2}\)
(see Eq.~(24) of Ref.~\cite{BPW1}).  Thus $B$ is fully calculable
within the model once the parameters $\sigma$, $\eta$, $\epsilon$,
and $y$ are determined in terms of inputs involving some quark and
lepton masses (as noted below).  In this way, one obtains $B
\approx (2.9 \pm 0.5)$.  The Dirac mass of the tau-neutrino is
obtained by using the $SU(4)$-color relation (see Eq.~(\ref{eq:1})):
\(m(\nu_{\mathrm{Dirac}}^{\tau}) \approx m_{\mathrm{top}}(M_{X})
\approx 120\ \mathrm{GeV}\). One thus obtains from Eq.~(12)
(as noted in Sec.~2):
\begin{eqnarray}
m(\nu_{3}) & \approx &
\frac{(2.9)(120 \mbox{\ GeV})^{2}}{10^{15} \mbox{\ GeV}} (1/2\mbox{--}2)
\nonumber \\
& \approx & (1/24 \mbox{\ eV})(1/2\mbox{--}2)
\label{seesaw2}
\end{eqnarray}
Noting that for hierarchical entries --- i.e. for ($\sigma$, $\epsilon$,
and $y$) \(\sim 1/10\) --- one naturally obtains a hierarchical spectrum
of neutrino-masses:
\(m(\nu_{1}) \lsim m(\nu_{2}) \sim (1/10)m(\nu_{3})\),
we thus get:
\begin{equation}
\left[ \sqrt{\Delta m_{23}^{2}} \right]_{\mathrm{Th}}
\approx m(\nu_{3}) \approx (1/24 \mbox{\ eV})(1/2\mbox{--}2)
\label{delta}
\end{equation}
This agrees remarkably well with the SuperK value of
\((\sqrt{\Delta m_{23}^{2}})_{\mathrm{SK}} (\approx 1/20 \mbox{
eV})\), which lies in the range of nearly (1/15 to 1/30) eV.  As
mentioned in the introduction, the success of this prediction
provides clear support for (i) the existence of $\nu_{R}$, (ii)
the notion of $SU(4)$-color symmetry that gives
$m(\nu_{\mathrm{Dirac}}^{\tau})$, (iii) the SUSY unification-scale
that gives $M_{R}$, \emph{and} (iv) the seesaw mechanism.

We note that alternative symmetries such as $SU(5)$ would have no
compelling reason to introduce the $\nu_{R}$'s.  Even if one did
introduce $\nu_{R}^{i}$ by hand, there would be no symmetry to relate the
Dirac mass of $\nu_{\tau}$ to the top quark mass.  Thus
$m(\nu_{\mathrm{Dirac}}^{\tau})$ would be an arbitrary parameter in $SU(5)$,
which, as noted in footnote~5, could well vary from say 1~GeV to 100~GeV.
Furthermore, without B-L as a local symmetry, the Majorana masses of the
RH neutrinos, which are singlets of $SU(5)$, can well be as high as the
string scale \(\sim 4 \times 10^{17}\ \mathrm{GeV}\) (say), and as low as
say 1~TeV.  Thus, as noted in footnote~5, within $SU(5)$, the absolute scale
of the mass of $\nu_3$, obtained via the familiar seesaw
mechanism \cite{seesaw},  would be uncertain by more than ten orders of
magnitude.

Other effective symmetries such as $[SU(3)]^{3}]$ \cite{su33}
and \(SU(2)_{L} \times SU(2)_{R}\times U(1)_{B-L}\times SU(3)^{C}\) \cite{2213}
would give $\nu_{R}$ and B-L as a local symmetry, but not the desired
$SU(4)$-color mass-relations: \(m(\nu_{\mathrm{Dirac}}^{\tau})
\approx m_{t}(M_{X})\) and \(m_{b}(M_{X}) \approx m_{\tau}\). Flipped
\(SU(5) \times U(1)\) \cite{flip} on the other hand would yield the desired
features for the neutrino-system, but not the
empirically favored $b$-$\tau$ mass
relation (Eq.~(1)). Thus, combined with the observed $b/\tau$ mass-ratio,
the SuperK data on atmospheric neutrino oscillation seems to
clearly select out the effective symmetry in 4D being either
$G(224)$ or $SO(10)$, as opposed to the other alternatives
mentioned above. \emph{It is in this sense that the neutrinos,
by virtue of their tiny masses, provide crucial information
on the unification-scale as
well as on the nature of the unification-symmetry in 4D, as
alluded to in the introduction}.

Ignoring possible phases in the parameters and thus the source of
CP violation for a moment, and also setting
$\zeta_{22}^d=\zeta_{22}^u=0$, as was done in Ref. \cite{BPW1},
the parameters ($\sigma$, $\eta$, $\epsilon$, $\epsilon'$,
$\eta'$, $\mathcal{M}_u^0$, $\mathcal{M}_D^0$, $y$) can be
determined by using, for example, \(m_t^{\mathrm{phys}} = 174
\mbox{\ GeV}\), \(m_c(m_c) =1.37 \mbox{\ GeV}\), \(m_s(1\mbox{\
GeV}) = 110\mbox{--}116 \mbox{\ MeV}\), \(m_u(1\mbox{ GeV})=6
\mbox{\ MeV}\), the observed masses of $e$, $\mu$, and $\tau$ and
$m(\nu_2)/m(\nu_3)\approx 1/(6\pm1)$ (as suggested by a
combination of atmospheric and solar neutrino data, the latter
corresponding to the LMA MSW solution, see below) as inputs. One
is thus led, {\it for this CP conserving case}, to the
 following fit for the parameters, and the
associated predictions \cite{BPW1}. [In this fit, we leave the
small quantities $x$ and $z$ in $M_R^\nu$ undetermined and proceed
by assuming that they have the magnitudes suggested by flavor
symmetries (i.e., \(x \sim (10^{-4}\mbox{--}10^{-5})\) and \(z
\sim (1/200)(1 \mbox{ to } 1/2)\) (see remarks below
Eq.~(\ref{eq:MajMM})]:
\begin{subequations}
\label{eq:fit}
\begin{eqnarray}
\sigma & \approx & 0.110 \\
\eta & \approx & 0.151 \\
\epsilon & \approx & -0.095\\
|\eta'| & \approx & 4.4 \times 10^{-3} \\
\epsilon' & \approx & 2\times 10^{-4} \\
\mathcal{M}_u^0 \approx m_t(M_X) & \approx & 120 \mbox{\ GeV} \\
\mathcal{M}^0_D \approx m_b(M_X) & \approx & 1.5 \mbox{\ GeV} \\
y & \approx & -1/17.
\end{eqnarray}
\end{subequations}
These output parameters remain stable to within 10\%
corresponding to small variations ($\lsim 10$\%)
in the input parameters of $m_{t}$, $m_{c}$, $m_{s}$, and $m_{u}$.
These in turn lead to the following predictions for the quarks and
light neutrinos \cite{BPW1,JCPErice}:
\begin{eqnarray}
\label{eq:pred}
\begin{array}{l}
m_b(m_b) \approx (4.7\mbox{--}4.9) \mbox{\ GeV},\\
\sqrt{\Delta m_{23}^2} \approx m(\nu_3) \approx \mbox{(1/24 eV)(1/2--2)},\\
\begin{array}{lcl}
V_{cb} & \approx & \left|\sqrt{\frac{m_s}{m_b}\left|\frac{\eta+\epsilon}
{\eta-\epsilon}\right|} - \sqrt{\frac{m_c}{m_t}\left|\frac{\sigma
+\epsilon}{\sigma-\epsilon}\right|}\right| \\
 & \approx & 0.044,
\end{array}\\
\left\{ \begin{array}{lcl}
\theta^{\mathrm{osc}}_{\nu_{\mu}\nu_{\tau}} & \approx &
\left|\sqrt{\frac{m_\mu}{m_\tau}}
\left| \frac{\eta-3\epsilon}{\eta+3\epsilon} \right|^{1/2} +
\sqrt{\frac{m_{\nu_2}}{m_{\nu_3}}}\right| \\
& \approx & |0.437+(0.378\pm 0.03)|,\\
\multicolumn{3}{l}{\mbox{Thus, } \sin^2 2\theta^{\mathrm{osc}}_{\nu_{\mu}\nu_{\tau}}\approx 0.993,} \\
\multicolumn{3}{l}{\mbox{ (for $\frac{m(\nu_2)}{m(\nu_3)}\approx 1/6$),}}\\
\end{array}\right.\\
V_{us}\approx \left|\sqrt{\frac{m_d}{m_s}}-\sqrt{\frac{m_u}{m_c}}\right|
\approx 0.20,\\
\left|\frac{V_{ub}}{V_{cb}} \right|\approx \sqrt{\frac{m_u}{m_c}}\approx
0.07,\\
m_d(\mbox{1 GeV})\approx \mbox{8 MeV}.
\end{array}
\end{eqnarray}
It is rather striking that all seven predictions in
Eq.~(\ref{eq:pred}) agree with observations, to within 10\%.
Particularly intriguing is the (B-L)-dependent {\it
group-theoretic correlation} between the contribution from the
first term in $V_{cb}$ and that in
$\theta^{\mathrm{osc}}_{\nu_{\mu}\nu_{\tau}}$, which explains
simultaneously why one is small ($V_{cb}$) and the other is large
($\theta^{\mathrm{osc}}_{\nu_{\mu}\nu_{\tau}}$) \cite{newFN36}.
That in turn provides some degree of confidence in the pattern of
the mass-matrices.

The Majorana masses of the RH neutrinos ($N_{iR}\equiv N_i$) are given by
\cite{JCPErice}:
\begin{eqnarray}
\label{eq:MajM}
M_{3}& \approx & M_R\approx 10^{15}\mbox{ GeV (1/2-1)},\nonumber\\
M_{2}& \approx & |y^2|M_{3}\approx \mbox{(2.5$\times 10^{12}$ GeV)(1/2-1)},\\
M_{1}& \approx & |x-z^2|M_{3} \sim (1/2\mbox{-}2)10^{-5}M_{3} \nonumber \\
 & & \sim 10^{10} \mbox{\ GeV}(1/4\mbox{--}2).\nonumber
\end{eqnarray}

Note that we necessarily have a hierarchical pattern for the light as well as
the heavy neutrinos (see discussions below on $m_{\nu_1}$).

As regards $\nu_e$-$\nu_{\mu}$ and $\nu_e$-$\nu_{\tau}$ oscillations, the
standard seesaw mechanism would typically lead to rather small angles
(e.g.\
\(\theta_{\nu_{e}\nu_{\mu}}^{\mathrm{osc}} \approx \sqrt{m_{e}/m_{\mu}}
\approx 0.06\)),
within the framework presented above \cite{BPW1}.
It has, however, been noted recently \cite{JCPErice} that small intrinsic
(non-seesaw) masses $\sim 10^{-3}$ eV of the LH neutrinos can arise quite
plausibly through higher dimensional operators of the form \cite{FN32}:
\(W_{12}\supset \kappa_{12}\mathbf{16}_1\mathbf{16}_2\mathbf{16}_H\mathbf{16}_H\mathbf{10}_H
\mathbf{10}_H/M_{\mathrm{eff}}^3\),
without involving the standard seesaw mechanism \cite{seesaw}.
One can verify that such a term would lead to an intrinsic Majorana mixing
mass term of the form $m_{12}^0\nu_L^e\nu_L^\mu$, with a strength given by
\(m_{12}^0\approx \kappa_{12}\langle\mathbf{16}_H \rangle^2(175\mbox{\ GeV})^2/
M_{\mathrm{eff}}^3\sim (1.5\mbox{--}6)\times 10^{-3}\) eV, for
$\langle\mathbf{16}_H \rangle\approx (1\mbox{-}2)M_{\mathrm{GUT}}$ and
\(\kappa_{12}\sim 1\),
if \(M_{\mathrm{eff}}\sim M_{\mathrm{GUT}} \approx
2\times 10^{16} \mbox{\ GeV}\) \cite{FN33}.
Such an intrinsic Majorana $\nu_e\nu_{\mu}$ mixing mass
$\sim $ few$\times 10^{-3}$ eV, though small compared to $m(\nu_3)$, is
still much larger than what one would generically get for the corresponding
term from the standard seesaw mechanism [as in Ref.~\cite{BPW1}]. Now, the
diagonal ($\nu_{\mu}\nu_{\mu}$) mass-term, arising from standard seesaw can
naturally be $\sim$ (3-8)$\times 10^{-3}$ eV for $|y|\approx 1/20$-1/15, say
\cite{BPW1}. Thus, taking the net values of
\(m_{22}^0\approx (6-7)\times 10^{-3}\) eV,
$m_{12}^0\sim 3\times 10^{-3}$ eV  as above and
$m_{11}^0\ll 10^{-3}$ eV, which are all plausible, we obtain
\(m_{\nu_2}\approx(6-7)\times 10^{-3}\) eV,
\(m_{\nu_1}\sim \mbox{(1 to few)} \times 10^{-3}\) eV, so that
\(\Delta m^2_{12}\approx (3.6\mbox{-}5)\times 10^{-5} \mbox{ eV}^{2}\) and
\(\sin^{2} 2\theta_{\nu_{e}\nu{\mu}}^{\mathrm{osc}}
\approx 0.6\mbox{--}0.7\). These go
well with the LMA MSW solution of the solar neutrino puzzle.

Thus, \emph{the intrinsic non-seesaw contribution} to the Majorana
masses of the
LH neutrinos can possibly have the right magnitude for $\nu_e$-$\nu_{\mu}$
mixing so as to lead to the LMA solution within the G(224)/SO(10)-framework,
without upsetting the successes of the seven predictions
in Eq.~(\ref{eq:pred}). [In contrast to the near maximality
of the
$\nu_{\mu}$-$\nu_{\tau}$ oscillation angle,
however, which emerges as a compelling
prediction of the framework \cite{BPW1}, the LMA solution, as obtained above,
should, be regarded as a consistent possibility, rather than as a compelling
prediction, within this framework.]

It is worth noting at this point that in a theory leading to Majorana
masses of the LH neutrinos as above, \emph{one would of course expect the
neutrinoless double beta decay process (like $n+n\rightarrow ppe^-
e^-$), satisfying $|\Delta L|=2$ and $|\Delta B|=0$, to occur at
some level.} The search for this process is most important because it
directly tests a fundamental conservation law and can shed light
on the Majorana nature of the neutrinos, as well as on certain CP
violating phases in the neutrino-system (assuming that the process is
dominated by neutrino-exchange). The crucial parameter which
controls the strength of this process is given by $m_{ee} \equiv
|\sum_i m_{\nu_i} U_{e i}^2|$. With a non-seesaw contribution leading
to \(m_{\nu_1}\sim\mbox{ few }\times 10^{-3}\mbox{ eV}\),
\(m_{\nu_2}\approx\mbox 7\times 10^{-3}\mbox{ eV}\),
\(\sin^2 2\theta_{12} \approx 0.6-0.7\), and an expected value for
\(\sin \theta_{13} \sim m^{0}_{13} / m^{0}_{33} \sim (1-5)\times 10^{-3}
\mbox{ eV }/(5\times 10^{-2}\mbox{ eV }) \sim (0.02-0.1)\), one would
expect \(m_{ee}\approx (1-5)\times 10^{-3}\mbox{ eV}\). Such a strength,
though compatible with current limits \cite{Vogel},
would be accessible if the current
sensitivity is improved by about a factor of 50--100. Improving the
sensitivity to this level would certainly be most desirable.

In summary, given the bizarre pattern of masses and mixings of the quarks,
charged leptons, and neutrinos, it seems truly remarkable that the simple
pattern of fermion mass matrices, motivated in large part by the group theory
of the $G(224)$ or $SO(10)$ symmetry and the minimality of the Higgs system,
and in part by the assumption of flavor symmetry
(of the type defined in Eq.~(10)),\footnote{One still needs
to understand the origin of flavor symmetries, for example, of the type
proposed here, in the context of the ground state solution of an underlying
theory like the string/M theory.}
leads to seven predictions in agreement with observations.  Particularly
significant are the predictions for $m(\nu_{L}^{3})$ (to within a factor
of 2 or 3, say), together with that of $m_b/m_{\tau}$, which help select out
the route to higher unification based on $G(224)$ or $SO(10)$ as the effective
symmetry in 4D, as opposed to other alternatives.  So also are the predictions
for the extreme smallness of $V_{cb}$ together with the near maximality of
$\theta^{\mathrm{osc}}_{\nu_{\mu}\nu_{\tau}}$.  I now proceed to present
briefly, in the next two sections, the results of some recent works on CP
and flavor violations and on baryogenesis via leptogenesis, all treated within
the same framework as presented here.

\section{CP and Flavor Violations Within the SUSY $G(224)/SO(10)$-Framework}
In this section I will present briefly some recent works by K.\ S.\ Babu,
Parul Rastogi, and myself which will appear in the form of two papers
\cite{Parul1,Parul2}.  At the outset I need to say a few words about
the origin of CP
violation within the G(224)/SO(10)-framework presented above. The discussion
so far
has ignored, for the sake of simplicity, possible CP violating phases in the
parameters ($\sigma$, $\eta$, $\epsilon$, $\eta'$, $\epsilon'$,
$\zeta_{22}^{u,d}$, $y$, $z$, and $x$) of the Dirac and Majorana mass matrices
[Eqs.~(\ref{eq:mat}, and (\ref{eq:MajMM})]. In general, however, these
parameters can and generically will have phases \cite{FN34}. Some combinations
of these phases enter into the CKM matrix and define the Wolfenstein parameters
$\rho_W$ and $\eta_W$, which in turn induce CP violation by
utilizing the standard model interactions.

Our procedure for dealing with CP and flavor violations may be summarized by
the following set of considerations:
\begin{enumerate}
\item Since the model is supersymmetric, CP and flavor violations naturally
arise also through s-fermion/gaugino loops involving scalar (mass)$^2$-
transitions which can preserve as well as flip chirality, such as
$(\tilde{q}^{i}_{L,R} \rightarrow \tilde{q}^{j}_{L,R})_{i \neq j}$ and
$(\tilde{q}^{i}_{L} \rightarrow \tilde{q}^{j}_{R})_{i \neq j}$ respectively.
These transitions (including their phases) get determined within
the model in terms of the fermion mass-matrices and the SUSY-parameters as
follows.
\item We assume that SUSY breaks at high scale $M^{*}(\sim M_{st}) \gsim
M_{GUT}$ such that the soft parameters are flavor-blind-and thus they are
family-universal at the scale $M^{*}$. A number of well-motivated models of
SUSY-breaking \cite{51Raby}-e.g. those based on the ideas of msugra,
gaugino-meditation, anomaly-meditation, dilaton-meditation or family-universal
anomalous U(1) D-term contribution or (preferably) on a combination of some
of these mechanisms-can induce such a breaking. In an extreme
version (such as CMSSM) such a universal model would involve only five
parameters ($m_0$, $m_{1/2}$, $a_0$, $\tan{\beta}$ and $\sgn(\mu)$), and in
some cases (as in \cite{FaraggiJCP}) $a_0$ would be zero at $M^*$. While for
most purposes we will use this restricted version of SUSY-breaking, including
$a_0=0$, as a guide,
we will not insist for example on Higgs-squark universality.
\item Although the squarks and sleptons of the three families have a common
mass $m_0$ and the off-diagonal (mass)$^2$-transitions (such as
$\tilde{b} \rightarrow \tilde{s}$ etc.) vanish at the scale $M^*$, SUSY
flavor violations arise in
the model as follows. Owing to flavor-dependent Yukawa couplings, with
$h_{top}$ being dominant, renormalization group running from $M^*$ to $M_{GUT}$
in the context of $SO(10)$ or $G(224)$ makes $\tilde{b}_{L,R}$ and
$\tilde{t}_{L,R}$ lighter than $(\tilde{d}, \tilde{s})_{L,R}$ \cite{52HBS}.

Now, following common practice, we analyze SUSY-contributions in the so-called
SUSY-basis, in which gluino-interactions are flavor-diagonal, with the quarks
being in their physical or mass basis (likewise for leptons). Let the quark
mass-matrices, defined originally in the gauge-basis, be diagonalized by the
matrices $X^{(q)}_{L,R}$ at the GUT-scale (where $q=u$ or $d$) so that the
CKM-matrix has the Wolfenstein-form. The squark (mass)$^2$ matrices
$M^{(0}_{LL}$ and $M^{(0)}_{RR}$, also defined in the gauge basis, must then be
transformed by the same matrices to the forms $X^{(q)\dagger}_L M^{(0)}_{LL}
X^{(q)}_L$, and likewise for $L \rightarrow R$.

Since $M^{(0)}_{LL}$ and $M^{(0)}_{RR}$, though diagonal, are not proportional
to unit matrices at the GUT-scale (for reasons explained above), their
transformations to the SUSY-basis (as above) would then induce flavor-violating
transitions such as $\tilde{b}_{L,R} \rightarrow \tilde{d}_{L,R}$, or
$\tilde{s}_{L,R}$ and $\tilde{d}_{L,R} \rightarrow \tilde{s}_{L,R}$ etc.,
that too with phases, depending upon $X^{(q)}_{L,R}$.
\emph{Note that these phases
and the associated CP violation
arise entirely from the quark mass-matrices, as also the CKM CP violation.}
\item Additional flavor-violations arise through RG-running of the
$\tilde{b}_L$-mass from $M_{GUT}$ to the electroweak scale in the context of
MSSM involving the top Yukawa coupling.
\item Furthermore, even if we start with $a_0=0$ at the high scale $M^*$, RG
running from $M^*$ to $M_{GUT}$ in the context of $SO(10)$ or $G(224)$ still
induces the A-parameters at the GUT-scale utilizing the gaugino-masses and the
Yukawa couplings \cite{52HBS}. These get determined within the model as well,
for a given choice of $m_0$, $m_{1/2}$ and $M^*/M_{GUT}$. These A-terms
induce chirality-flipping transitions such as
$\tilde{s}_L \rightarrow \tilde{d}_R$, $\tilde{b}_L \rightarrow \tilde{s}_R$,
$\tilde{\mu}_L \rightarrow \tilde{e}_R$, $\tilde{d}_L \rightarrow \tilde{d}_R$
and $\tilde{e}_L \rightarrow \tilde{e}_R$, which can be important for
$\epsilon'_K$, $B_d \rightarrow \Phi K_S$, $\mu \rightarrow e \gamma$ and
edm's of the neutron and the electron.
\item \emph{The interesting point is that the net values of the off-diagonal
squark-mixings including their phases, and thereby the flavor and CP violations
induced by them, are entirely determined within our approach by the entries
in the quark mass-matrices and the choice of ($m_0$, $m_{1/2}$, $a_0$,
$\tan{\beta}$ and $\sgn(\mu)$); similarly for the leptonic sector.} Within the
$G(224)/SO(10)$ framework presented in sec.~4, the quark mass-matrices are
however tightly constrained by our considerations of fermion masses and
neutrino-oscillations. This is the reason why, within our approach, \emph{SUSY
CP and flavor violations get intimately linked with fermion masses and
neutrino-oscillations} \cite{53Babu,54Chang}.
\end{enumerate}

With this to serve as a background, since we wish to introduce CP violation by
introducing phases into at least some of the entries in the quark (and
thereby lepton) mass-matrices, the question arises:
%\fbox{\parbox{\columnwidth}{}}

Can observed CP and/or flavor-violations in the
quark and lepton sectors
(including the empirical limits in some of these) emerge consistently within
the $G(224)/SO(10)$-framework, for \emph{any} choice of phases in the fermion
mass-matrices of Eq.~(\ref{eq:mat}), while preserving all its successes
with respect to fermion masses and neutrino oscillations?

This is indeed a \emph{non-trivial challenge} to meet within the $SO(10)$ or
$G(224)$-framework, since  the constraints from both CP and flavor violations
on the one hand and fermion masses on the other hand are severe.

Turning to experiments, there are now four well-measured entities reflecting
flavor and/or CP violations in the quark-sector which confront theoretical
ideas on physics beyond the standard model.  They are:\footnote{$\epsilon'_K$
reflecting direct CP violation is well-measured, but its theoretical
implications are at present unclear  due to uncertainties in the matrix
element}
\begin{equation}
\Delta m_K, \epsilon_K, \Delta m_{B_d}\mbox{ and }
S(B_d \rightarrow J/\Psi K_s).
\label{eq:20}
\end{equation}
It is indeed remarkable that the observed values including the signs of
\emph{all four} entries as well as the lower limit on $\Delta m_{B_s}$ can
consistently be realized (allowing for uncertainties in matrix elements of
up to 20\%) within the standard CKM-model for a single choice of the
Wolfenstein parameters \cite{55Ciuccini}:
\begin{equation}
\bar\rho_W = 0.178 \pm .046;\; \bar\eta_W=0.341 \pm .028.
\label{eq:21}
\end{equation}
In particular, using the observed values of $\epsilon_K=2.27 \times 10^{-3}$,
$|V_{ub}|=(3.55 \pm .36)\times 10^{-3}$, $|V_{cb}|= 4.1\pm .16)\times 10^{-2}$,
$\Delta m_{B_d}=(3.3 \pm .06) \times 10^{-13}$ GeV, and the upper limit on
$\Delta m_{B_d}/\Delta m_{B_s} >0.035$, one can phenomenologically determine
$\bar{\rho_W}$ and $\bar{\eta_W}$ in the SM and predict the asymmetry parameter
$S(B_d \rightarrow J/\Psi K_s)$ to be $\approx 0.70 \pm 0.1$ \cite{56ASoni}.
This agrees remarkably well with the observed value
$S(B_d \rightarrow J/\Psi K_s)_{expt}=0.734 \pm 0.054$ \cite{57BaBarBELLE}.
This agreement of the SM in turn poses a challenge for physics beyond the SM
especially for SUSY GUT-models possessing CP and flavor-violations as
described above. The question is: If such a GUT model is constrained, as in
our case, by requiring that it should successfully describe the fermion and
neutrino oscillations, can it still yield $\bar\rho_W$ and $\bar\eta_W$ more or
less in accord with the values given above? In particular, adding contributions
from the standard model interactions as well as from SUSY-graphs, can such a
constrained $SO(10)$ or $G(224)$ model account for the observed values of the
four entities listed in Eq.~(\ref{eq:20})?

First of all, one might have thought, given the freedom in the choice of
phases in the parameters of the mass-matrices, that it ought to be possible
to get $\bar{\rho}_W$ and $\bar{\eta}_W$ in accord with the SM-values given in
eq.~(\ref{eq:21}), within any $SO(10)$-model. It turns out, however, that in
general this is indeed not possible without running into a conflict with the
fermion masses and/or neutrino-oscillation parameters\footnote{For a discussion
of the difficultities in this regard within a recently proposed $SO(10)$-model,
see e.g. Ref.~\cite{58GohRNM}.}.
In other words, any predictive $SO(10)$-model is rather constrained in this
regard.

Second, one might think that even if the derived values of $\bar\rho_W$ and
$\bar\eta_W$, constrained by the pattern of fermion masses and neutrino
oscillations, are found to be very different in signs and/or in magnitudes
from the SM-values shown in Eq.~(\ref{eq:21}), perhaps the SUSY-contributions
added  to the new SM-contributions (based on the derived values of
$\bar\rho_W$ and $\bar\eta_W$) could possibly account for all four
entities listed in Eq.~(\ref{eq:20}). \emph{It seems to us, however, that this
is simply not a viable and natural possibility.} This is because the SUSY
contributions
combine in different ways with the SM-contributions for the four different
entities listed in Eq. (\ref{eq:20}). Suppose the derived values of
$\bar\rho_W$ and $\bar\eta_W$ are very different from the values given in
Eq.~(\ref{eq:21}).
The SM-contributions to each of $\epsilon_K$, $\Delta m_{B_d}$ and
$S(B_d \rightarrow J/\Psi K_s)$\footnote{$\Delta m_K$ depends primarily on
$(V_{cs}V^*_{cd})^2$ and thus, to a good approximation, is independent of
$\bar\rho$ and $\bar\eta$} would in this case be in gross conflict with
observations. That means that the SUSY-contribution would have to be sizable
for each of these three entities, for this case, so as to possibly
compensate for errors in
the SM-contributions. Now, the magnitude of the SUSY-contribution can perhaps
be adjusted (by choosing e.g s-fermion mass) so that the combined
contribution from (SM+SUSY) would give the
right value for \emph{one} of the three entities, but it would be rather
impossible that the SM and SUSY contributions would add in just the right
way in sign
and magnitude so that the net contribution for the other two entities, as well
as that for $\Delta m_K$, would agree with observations.

Of course by introducing a completely arbitrary set of soft SUSY-breaking
parameters in the gauge-basis, including
intrinsic phases (in general there
are some $105$ parameters for MSSM), perhaps an agreement can be realized with
respect to all four entities listed in Eq.~(\ref{eq:20}), even if the derived
values of $\bar{\rho}_W$ and $\bar{\eta}_W$ are very different from the SM
values given in Eq.~(\ref{eq:21}). This would, however, be a rather unnatural
solution, with
many arbitrary parameters. And the question would still arise: If, in the true
picture SUSY-contribution is so important, why does the SM (with zero SUSY
contribution) provide such an excellent description for all four
entities with the right prediction for $S(B_d \rightarrow J/\Psi K_s)$ in the
first place?

This is why it seems to us that \emph{the only viable and natural solution}
for any SUSY $G(224)$ or
$SO(10)$-model for fermion masses and neutrino-oscillations is that, the model,
with allowance
for phases in the fermion mass-matrices, should not only yield the
masses and mixings of all fermions including neutrinos in accord with
observations (as in Sec.~4), but it should yield $\bar\rho_W$ and
$\bar\eta_W$ that are close to the values shown in Eq.~(\ref{eq:21}). This
would be a major step in the right direction. One then needs to ask: how does
the combined (SM + SUSY) contributions fare for such a solution as regards its
predictions for the four entities of Eq.~(\ref{eq:20}) and other CP and/or
flavor-violating processes for any given choice of SUSY parameters
($m_0$, $m_{1/2}$, $a_0$, $\tan{\beta}$ and $\sgn(\mu)$)?

Without further elaboration, I will now briefly summarize the main results of
Refs.~\cite{Parul1} and \cite{Parul2}. Some of these results and fittings
should be regarded as preliminary.
\begin{enumerate}
\item Allowing for phases ($\sim 1/10$ to $\sim 1/2$) in the parameters
$\eta$, $\sigma$,
$\epsilon'$ and $\zeta^d_{22}$ of the $G(224)/SO(10)$-framework
(see Eq.~(\ref{eq:mat})) we find that there do exist solutions which yield
masses
and mixings of quarks and leptons including neutrinos all in accord   with
observations (to within 10 \%), and at the same time yield the following values
for the Wolfenstein parameters:
\begin{equation}
\hat\rho_W\approx 0.17,\; \hat\eta_W \approx 0.31.
\label{eq:22}
\end{equation}
The hat here signifies that these are the
values of $\bar\rho_W$ and $\bar\eta_W$
which are derived (for a suitable choice of phases in the parameters as
mentioned above) from within the structure
of the mass-matrices (Eq.~(\ref{eq:mat})), subject to the constraints of
fermion masses and mixings, while the corresponding
phenomenological values are listed in Eq.~(\ref{eq:21}).

Note, as desired, the $G(224)/SO(10)$-framework presented here has turned out
to be capable
of yielding $\hat\rho_W$ and $\hat\eta_W$ close to the SM-values of
$\bar\rho_W$ and $\bar\eta_W$ while preserving the successes with respect to fermion
masses and neutrino oscillations as in Sec.~4. As mentioned above, this
is non-trivial. A priori a  given $SO(10)$-model of fermion masses
and neutrino-oscillations may not in fact be capable of yielding ($\hat\rho_W,
\hat\eta_W$) even lying in the first quadrant, for any choice of phases of
the relevant parameters. The difficulty with such a situation,
should it emerge, is mentioned above.
\item Including both the SM-contribution (with $\hat\rho_W$
and $\hat\eta_W$ as above) and the SUSY-contribution (with a plausible choice
of the spectrum-e.g. $m_{sq}\approx (0.8-1)$ TeV and
$x=(m^2_{\tilde{g}}/m^2_{sq})\approx 0.6-0.8$), we obtain \cite{Parul1}:
\begin{eqnarray}
(\Delta m_K)_{short dist}\approx 3\times 10^{-15}\mbox{ GeV};\nonumber\\
\epsilon_K \approx(2\, \mbox{to}\, 2.4)\times 10^{-3};\nonumber\\
\Delta m_{B_d} \approx (3\, \mbox{to}\, 3.3) \times 10^{-13}\mbox{ GeV};
\nonumber\\
S(B_d \rightarrow J/\Psi K_s) \approx 0.65 - 0.68.
\label{eq:23}
\end{eqnarray}

We have used $\hat{B}_k=0.87$ and $f_{Bd}\sqrt{\hat{B}_{Bd}}=200$
MeV (see \cite{55Ciuccini}). Now the first four on which there is
reliable data are in good agreement with observations (within
10\%). The spectrum of ($m_{sq}$, $m_{\tilde{g}}$) considered
above can be realized, for example for a choice of $(m_0,
m_{1/2})\approx (600, 220)$ GeV. Other choices of
SUSY-parameters-for example $(m_0, m_{1/2})= (60, 260)$ GeV, or
$(100, 440)$ GeV, or $(1000, 250)$ GeV --- which would be in
accord with the WMAP-constraint \cite{59EOS} in the event that the
LSP is the cold dark matter (see remarks later), also lead to
quite acceptable values for all four entities listed above
\cite{Parul1}. In all these cases, the SUSY-contribution turns out
to be rather small ($\lsim 5 \%$ in amplitude), except however for
$\epsilon_K$, for which it is sizable ($\approx 20 - 30 \%$) and
has opposite sign, compared to the SM-contribution.

\emph{We thus see that the SUSY $G(224)$ or $SO(10)$-framework (remarkably
enough) has met the challenges so far in being able to reproduce the observed
features of both CP and quark-flavor violations as well as fermion masses and
neutrino-oscillations!}
Owing to introduction of four phases, the number of parameters has increased
compared to that in Sec.~4, but the number of observable entities involving
CP and flavor violations including those in the $B_s$ and lepton-systems, only
some of which will be presented here, has increased many-fold. Thus the
framework will be thoroughly testable as regards its predictions for CP and
flavor-violations, especially once the SUSY-parameters are determined by
(hopefully successful) SUSY-searches.

\item As noted in \cite{53Babu,54Chang}, the mass-parameter
$\delta^{23}_{RR}(\tilde{b}_R\rightarrow \tilde{s}_R)$ gets enhanced both due
to (a)~the SUSY flavor-violation arising from RGE running from $M^*$ to
$M_{GUT}$ in the context of $SO(10)$, and equally important (b)~large
$\nu_{\mu}-\nu_{\tau}$ oscillation angle. This enhancement is found to be
insufficient, however, within our model, to produce a large deviation in
$S(B_d \rightarrow \Phi K_s)$ from the SM-prediction. This situation is
found to prevail even after the inclusion of the chirality-flipping A-term
contribution to $\delta^{23}_{RL}(\tilde{b}_R\rightarrow \tilde{s}_L)$, where
the A-term is induced (as mentioned in (\ref{eq:scheme})) trough RG running
from $M^*$ to $M_{GUT}$. As a result, we obtain:
\begin{equation}
S(B_d \rightarrow \Phi K_s) \approx 0.65.
\label{eq:24}
\end{equation}
Thus the framework predicts that $S(B_d \rightarrow \Phi K_s)$ will be close
to the SM-prediction ($\approx 0.70\pm .10$) and certainly not negative in
sign. At present BaBar and BELL data yield widely differing values of
($0.45 \pm 0.43 \pm 0.07$) and ($-0.96\pm 0.50^{+0.09}_{-0.07}$) respectively
for $S(B_d \rightarrow \Phi K_s)$ \cite{57BaBarBELLE}. It will thus be
extremely interesting from the viewpoint of the $G(224)/SO(10)$-framework
presented here to see whether the true value of $S(B_d \rightarrow \Phi K_s)$
will turn out to be close to the SM-prediction or not.

\item \textbf{Lepton Flavor Violations}: As regards lepton flavor-violations
($\mu\rightarrow e\gamma,\, \tau \rightarrow\mu\gamma$ etc.) we get
contributions from \emph{three sources}: (i) $(\delta m^2)^{ij}_{LL}$ arising
from
RG-running from $M^*$ to $M_{GUT}$, in the context of  $SO(10)$ or $G(224)$,
involving the large top Yukawa coupling; (ii) $(\delta \acute{m}^2)^{ij}_{LL}$
arising from RG-running from $M_{GUT}$ to the RH neutrino mass-scales
$M_{Ri}$ involving $\nu_R^i$ Yukawa-couplings (corresponding to
Eq.~(\ref{eq:mat})); and (iii) chirality-flipping $(\delta m^2)^{ij}_{LR}$
arising from A-terms, induced through RG-running from $M^*$ to $M_{GUT}$ in
the
context of $SO(10)$ or $G(224)$, involving gaugino-masses and Yukawa couplings.
Note that all three contributions are tied to our fermion mass-matrices
including those of the neutrinos whose successes are discussed in Sec.~4.
They are thus fixed in our model for a given choice of $m_0$, $m_{1/2}$,
and $M^*/M_{GUT}$.
\end{enumerate}

There is a vast literature on the subject of lepton flavor violation (LFV).
(For earlier works see Ref.~\cite{60BM}; and for a partial list of references
including recent works see Ref.~\cite{61MVV}). Most of the works in the
literature have focused
on the contribution from the second source (involving the Yukawa couplings of
the RH neutrinos) which is proportional to $\tan\beta$ in the amplitude.
It turns out, however, that the contribution from the first source
$(\delta m^2)^{ij}_{LL}$ arising from $SO(10)$-running from $M^*$ to $M_{GUT}$
(which is proportional to $\tan\beta$) and that from the third source arising
from the induced A-terms ($\propto 1/\tan\beta$) are in fact the dominant ones
for $\tan\beta\lsim1 0$,
as long as $\ln(M^*/M_{GUT})\gsim 1$. We consider the contribution from all
three sources by summing the corresponding amplitudes, and by varying
($m_0$, $m_{1/2}$, $\tan{\beta}$ and $\sgn(\mu)$). Details of these results
will appear in Ref.~\cite{Parul2}. Here I present the results for a fixed
value of $\tan{\beta}=10,\ln(M^*/M_{GUT})=1$, and four sample choices for
$(m_0, m_{1/2})=(500, 200)$ GeV (Case I), $(700, 300)$ GeV (Case II),
$(100, 440)$ GeV (Case III), and $(1000, 250)$ GeV (Case IV):
\begin{equation}
\begin{array}{ccc}
& B(\mu\rightarrow e \gamma) & B(\tau\rightarrow\mu \gamma) \nonumber \\
& (\mu > 0, \mu < 0) & (\mu > 0, \mu > 0) \nonumber \\
I & (0.4,\, 1.9)\times 10^{-11} & (7.4,\, 8.4) \times 10^{-9} \nonumber\\
II & (1.3,\, 5.5)\times 10^{-12} & (1.7,\, 2.0) \times 10^{-9}\nonumber\\
III & (1.4,\, 1.4)\times 10^{-8} & (8.8,\, 9.0) \times 10^{-8}\nonumber\\
IV & (2.1,\, 7.4)\times 10^{-13} & (8.6,\, 8.2) \times 10^{-10}
\end{array}
\label{eqn:25}
\end{equation}
Of the four cases exhibited above, Case III
(low $m_0\approx 100$ GeV, with high $m_{1/2} \sim 4.4 m_0$) and Case IV (high
$m_0\sim$ 1 TeV, with low $m_{1/2}$) are in accord with the WMAP-constraint,
assuming that the lightest neutralino is the LSP and it represents cold dark
matter (see e.g. Ref.~\cite{59EOS} for a recent analysis that allows for
uncertainties in ($m_t$, $m_b$)), while Cases I and II per say are not,
\emph{if} we stick to the assumption stated above. It seems to me that Cases
like I and II can well be consistent with the WMAP-data, however, under a
variety of circumstances including the possibility that R-parity is broken
mildly say by a bilinear term $(\kappa L H_u)$ in the superpotential so that
the lightest neutralino
(LSP) decays with a lifetime $\sim 10^{-4} - 10^{-5}$ sec, say, to
ordinary particles long before nucleosynthesis \cite{62CDE}, and that some
other particle like the axion provides CDM. An alternative possibility,
considered for example in \cite{63BS}, is that the axino is the LSP and
provides cold dark matter (subject to R-parity conservation). I do not wish
to enter into a detailed discussion of this issue here, except to say that it
seems prudent to keep an open mind at present as regards all four choices of
($m_0$, $m_{1/2}$) exhibited above, and their variants, and study their
phenomenological consequences. Given the current
limits of $B(\mu\rightarrow e \gamma) \leq 1.2 \times 10^{-11}$
\cite{64Brooks} and $B(\tau\rightarrow\mu \gamma) \leq 3.11.2 \times 10^{-7}$
\cite{65Abe}, we see that while Case III (low $m_0\approx 100$ GeV, and high
$m_{1/2} \sim 4.4 m_0$) is clearly excluded\footnote{The sharp increase in
$B(\mu\rightarrow e\gamma)$ for Case III (low $m_0$ and high $m_{1/2}$) is
entirely because of the strong enhancement of the induced A-term contribution
which is very roughly proportional to
$(m_{1/2})^2/(m^2_{\tilde{l}L} m^2_{\tilde{l}R})$, in the
amplitude. Thus it increases sharply both because of large $m_{1/2}$ \emph{and}
small $m_0$ for case III. As noted above, this induced A-term contribution
which arises from $SO(10)$-running from$M^*$ to $M_{GUT}$ has invariably been
omitted in the literature. It should, however, be present in any SUSY $SO(10)$
or $G(224)$-model, if $M^* > M_{GUT}$.} by the limit on
$\mu\rightarrow e\gamma$-decay (taking $\ln M^*/M_{GUT}\geq 1$), the other
three cases are fully compatible with the present limits. But they clearly
imply that $\mu\rightarrow e\gamma$-decay should be observed with an
improvement in the current limit by a factor of $10-100$. Thus the
$G(224)/SO(10)$-framework for fermion masses,
neutrino oscillations and CP-violation \cite{Parul1} presented here, will have
its stringent tests once the current limit especially on the branching ratio
for $\mu\rightarrow e \gamma$-decay is improved by such a factor.

\textbf{Electric Dipole Moments in the $G(224)/SO(10)$-Model}

As regards CP violation, another important prediction of the model is on the
edm's for the neutron and the electron. As mentioned above, RG running from
$M^*$ to $M_{GUT}$ induces the A-term which in turn generates
chirality-flipping transitions such as $\tilde{q}_L\rightarrow\tilde{q}_R$
and $\tilde{l}_L\rightarrow\tilde{l}_R$. These having phases through the
fermion mass-matrices induce edm's. Given the fermion mass-matrices as in
Sec.~4 and the phases determined by our analysis of CP and flavor violations
in the quark-system (as discussed in the beginning of this section), all the
relevant s-fermion mass-parameters --- i.e\. Im$(\delta^d_{LR})_{11}$,
Im$(\delta^u_{LR})_{11}$, and Im$(\delta^l_{LR})_{11}$-are completely known
within our model, for a given value of $M^*/M_{GUT}$ and that of $\tan\beta$.
These in turn allow us to predict the edm's for a given choice of $m_0$,
$m_{1/2}$, $\tan{\beta}$ and $M^*/M_{GUT}$. Details of the results as a
function of these SUSY-parameters will be presented in Ref.~\cite{Parul2}.
The predictions for a specific choice ($m_{\tilde q}=m_{\tilde g}=600$ GeV,
$\ln(M^*/M_{GUT})=1$) is given bellow \footnote{Intrinsic SUSY-phases
such as that in
$(\mu m^*_2)$, if present, can make additional contributions to edm's which
should be added to the contribution shown above. This contribution would
increase with $\tan\beta$. In a theory where such instrinsic phases are
naturally zero or small, these contributions can of course be dropped.}:
\begin{equation}
d_n=\left\{\begin{array}{ll}
5.8\times 10^{-26}\mbox{e cm} & (\tan\beta=5)\\
3.8\times 10^{-26}\mbox{e cm} & (\tan\beta=10)
\end{array}
\right.\label{eq:26}
\end{equation}
\begin{equation}
d_e=\frac{(2.67\times 10^{-28}\mbox{e cm})}{\tan\beta} \ \ (m_{\tilde l}=500\mbox{ GeV})
\label{eq:27}.
\end{equation}
Note that the A-term contribution is larger for smaller $\tan{\beta}$ (For
many reasons, including constraints from proton lifetime, small
$\tan{\beta}\leq 10$ is preferred).

Given the experimental limits $d_n < 6.3 \times 10^{-26}$e cm \cite{66Harris}
and $d_e < 4.3 \times 10^{-27}$e cm \cite{67Commins}, we see that the
predictions of the model especially for the edm of the neutron is in an
extremely interesting range suggesting that it should be discovered with an
improvement of the current limit by about a factor of $10$.

In summary for this section, we see that $G(224)/SO(10)$-framework provides a
phenomenologically viable and a unified picture of fermion masses, neutrino
oscillations as well as CP and flavor violations. One question on the framework
is that it does not provide (e.g. by symmetry-arguments including flavor
symmetries) any guidance on the phases-their magnitude and signs. This remains
a challenge for the future. On the positive side, the framework not only
provides a consistent and predictive picture as regards a vast set of phenomena
noted above but also presents several crucial tests including those on the
asymmetry parameter for $(B_d \rightarrow \Phi K_s)$, branching ratio for
$\mu\rightarrow e \gamma$-decay and edm's of the neutron.

I next discuss the issue of baryogenesis within the same framework.

\section{Baryogenesis Via Leptogenesis Within the $G(224)/SO(10)$-Framework}

The observed matter-antimatter asymmetry provides an important clue to physics
at truly short distances. Given the existence of RH neutrinos, as required by
the symmetry $SU(4)$-color or $SU(2)_R$, possessing superheavy Majorana masses
which violate B-L by two units, baryogenesis via leptogenesis \cite{FYKRS} has
emerged as perhaps the most viable and natural mechanism for generating the
baryon asymmetry of the universe. The most interesting aspect of this mechanism
is that it directly relates our understanding of the light neutrino masses to
our own origin. The question of whether this mechanism can quantitatively
explain the magnitude of the observed baryon-asymmetry depends however
crucially on the Dirac as well as the Majorana mass-matrices of the neutrinos,
including the phases and the eigenvalues of the latter-i.e. $M_1$, $M_2$ and
$M_3$ (see Eq.~(\ref{eq:MajM})).

This question has been considered in a recent work \cite{68JCP} in the context
of a realistic and predictive framework for fermion masses and neutrino
oscillations, based on the symmetry $G(224)$ or $SO(10)$ , as discussed in
Sec.~4, with CP violation treated as in Sec.~5. It has also been discussed in
a recent review \cite{JCPErice}. Here I will primarily quote the results and
refer the reader to Ref.~\cite{68JCP} for more details especially including
the discussion on inflation and relevant references.

The basic picture is this. Following inflation, the lightest RH neutrinos
($N_1$'s) with a mass $\approx 10^{10}$ GeV ($1/3\ -\ 3$) are produced either
from the thermal bath following reheating ($T_{RH}\approx \mbox{ few} \times
10^9$ GeV), or non-thermally directly from the decay of the inflaton
\footnote{In this case the inflaton can naturally be composed of the Higgs-like
objects having the quantum numbers of the RH sneutrinos ($\tilde{\nu}_{RH}$ and
$\tilde{\bar{\nu}}_{RH}$) lying in $(1,\ 2,\ 4)_H$ and $(1,\ 2,\ \bar{4})_H$
for $G(224)$ (or $16_H$ and $\bar{16}_H$ for $SO(10)$), whose VEV's break B-L
 and give Majorana masses to the RH neutrinos via the coupling shown in
Eq.~(\ref{eq:LMaj}).} (with $T_{RH}$ in this case being about $10^7\ -\ 10^8$
GeV). In either case, the RH neutrinos having Majorana masses decay by
utilizing their Dirac Yukawa couplings into both $l+H$ and $\bar{l}+\bar{H}$
(and corresponding SUSY modes), thus violating B-L. In the presence of CP
violating phases, these decays produce a net lepton-asymmetry
$Y_L=(n_L-n_{\bar{L}})/s$ which is converted to a baryon-asymmetry
$Y_B=(n_B-n_{\bar{B}})/s=C Y_L$ ($C\approx -1/3$ for MSSM) by the EW sphaleron
effects. Using the Dirac and the Majorana mass-matrices of Sec.~4, with the
introduction of CP-violating phases in them as discussed in Sec.~5, the
lepton-asymmetry produced per $N_1$ (or ($\tilde{N}_1+\bar{\tilde{N}}_1)$-pair)
decay is found to be \cite{68JCP}:
\begin{eqnarray}
\epsilon_{1} & \approx & \frac{1}{8\pi}\left(\frac{\mathcal{M}_u^0}{v}\right)^2
|(\sigma+3\epsilon)-y|^2\sin\left(2\phi_{21}\right)\nonumber\\
&    &  \times (-3)\left(\frac{M_1}{M_2}\right)\nonumber\\
& \approx & -\left(2.0\times 10^{-6}\right) \sin\left(2\phi_{21}\right)
\nonumber\\
&    &
\times\left[\frac{(M_1/M_2)}{5\times 10^{-3}}\right]
\label{eq:28}
\end{eqnarray}
\emph{Here $\phi_{21}$ denotes an effective phase depending upon phases in the
Dirac as well as Majorana mass-matrices (see Ref.~\cite{68JCP}).} Note that the
parameters $\sigma$, $\epsilon$, $y$ and $(\mathcal{M}_u^0/v)$ are already
determined within
our framework (to within 10 \%) from considerations of fermion masses and
neutrino oscillations (see Sec.~4 and 5). Furthermore, from Eq.~(\ref{eq:MajM})
we see that $M_1\approx(1/3 - 3)\times 10^{10}$ GeV, and
$M_2\sim 2\times10^{12}$ GeV, thus
$M_1/M_2\approx(5\times 10^{-3})(1/3 - 3)$. In short, leaving aside the phase
factor, the RHS of Eq.~(\ref{eq:28}) is pretty well determined within our
framework (to within about a factor of 5), as opposed to being uncertain by
orders of magnitude either way. \emph{This is the advantage of our obtaining
the lepton-asymmetry in conjunction with a predictive framework for fermion
masses and neutrino oscillations.}
Now the phase angle $\phi_{21}$ is uncertain because
we do not have any constraint yet on the phases in the Majorana sector
$(M^\nu_R)$. At the same time, since the phases in the Dirac sector are
relatively large (see Sec.~5 and Ref.~\cite{Parul1}), barring unnatural
cancelation between the Dirac and Majorana phases, we would naturally expect
$\sin(2\phi_{21})$ to be sizable-i.e. of order $1/10$ to $1$ (say).

The lepton-asymmetry is given by $Y_L= \kappa (\epsilon_1/g^*)$, where $\kappa$
denotes an efficiency factor representing wash-out effects and $g^*$ denotes
the light degrees of freedom ($g^*\approx 228$ for MSSM). For our model, using
recent discussions on $\kappa$ from Ref.~\cite{69BDB}, we obtain:
$\kappa\approx(1/18 - 1/60)$, for the thermal case, depending upon the $''31''$
entries in the neutrino-Dirac and Majorana mass-matrices (see
Ref.~\cite{68JCP}). Thus, for the thermal case, we obtain:
\begin{equation}
(Y_B)_{thermal}/\sin(2\phi_{21})\approx (10 - 30)\times 10^{-11}
\label{eq:29}
\end{equation}
where, for concreteness, we have chosen $M_1\approx 4\times 10^9$ GeV and
$M_2\approx 1\times 10^{12}$ GeV, in accord with Eq.~(\ref{eq:MajM}). In this
case, the reheat temperature would have to be about few $\times 10^9$ GeV so
that $N_1$'s can be produced thermally. We see that the derived values of
$Y_B$ can in fact account for the recently observed value
$(Y_B)_{WMAP}\approx (8.7 \pm 0.4)\times 10^{-11}$ \cite{70WMAP}, for a
natural value of the phase angle $\sin(2\phi_{21})\approx (1/3- 1)$. As
discussed
below, the case of non-thermal leptogenesis can allow even lower
values of the phase angle. It also typically yields a significantly lower
reheat temperature ($\sim10^7 - 10^8$ GeV) which may be in better accord with
the gravitino-constraint.

For the non-thermal case, to be specific one may assume an effective
superpotential \cite{71Shafi}:
$W^{infl}_{eff}=\lambda S (\bar{\Phi}\Phi-M^2) +$ (non-ren. terms) so as to
implement hybrid inflation; here $S$ is a singlet field and $\Phi$ and
$\bar{\Phi}$ are Higgs fields transforming as $(1, 2, 4)$ and $(1, 2, \bar{4})$
of $G(224)$ which break B-L at the GUT scale and give Majorana masses to the RH
neutrinos. Following the discussion in \cite{71Shafi,68JCP}, one obtains:
$m_{infl}=\sqrt 2 \lambda M$, where $M=<(1, 2, 4)_H>\approx 2\times 10^{16}$
GeV; $T_{RH}\approx(1/7)(\Gamma_{infl}M_{Pl})^{1/2}\approx(1/7)(M_1/M)
(m_{infl}M_{Pl}/8\pi)^{1/2}$ and $Y_B\approx-(1/2)(T_{RH}/m_{infl})
\varepsilon_1$. Taking the coupling $\lambda$ in a plausible range $(10^{-5} -
10^{-6})$ (which lead to the desired reheat temperature, see below) and the
asymmetry-parameter $\varepsilon_1$ for the
$G(224)/SO(10)$-framework as given in Eq.~(\ref{eq:28}), the baryon-asymmetry
$Y_B$ can then be derived. The values for $Y_B$ thus obtained are listed in
Table~\ref{tab:1}.
\begin{table}
\begin{tabular}{|l|l|l|} \hline
$\lambda$ &  $10^{-5}$ &  $10^{-6}$ \\ \hline \hline
$m_{infl}$ GeV &  $3\times 10^{11}$
    & $3\times 10^{10}$ \\ \hline
$T_{RH}$ GeV &  $(5.3-1.8)\times 10^7$ &
    $(17-5.6)\times 10^6$ \\ \hline
$\frac{Y_B\times 10^{11}}{\sin(2\phi_{21})}$ & $(100-10)$
    & $(300-33)$ \\ \hline
\end{tabular}
\caption{Baryon Asymmetry For Non-Thermal Leptogenesis}
\label{tab:1}
\end{table}

The variation in the entries correspond to taking $M_1=(2\times 10^{10}
\mbox{ GeV})(1-1/3)$ with  $M_2=(2\times 10^{12})$ GeV in accord with
Eq.~(\ref{eq:MajM}). We see that for this case of non-thermal leptogenesis,
one
quite plausibly obtains $Y_B\approx(8 - 9)\times10^{-11}$ in full accord with
the WMAP data, for natural values of the phase angle
$\sin(2\phi_{21})\approx(1/3 - 1/10)$, and with $T_{RH}$ being as low as $10^7$
GeV $(2-1/2)$. Such low values of the reheat temperature are fully consistent
with the gravitino-constraint for $m_{3/2}\approx 400$ GeV $- 1$ TeV (say),
even if one allows for possible hadronic decays of the gravitinos for example
via $\gamma\tilde{\gamma}$-modes \cite{72KKM}.

In summary, I have presented two alternative scenarios (thermal as
well as non-thermal) for inflation and leptogenesis. We see that
the $G(224)/SO(10)$-framework provides a simple and unified
description of not only fermion masses, neutrino oscillations
(consistent with maximal atmospheric and large solar oscillation
angles) \emph{and} CP violation, but also of baryogenesis via
leptogenesis, in either scenario. Each of the following features -
(a)~the existence of the RH neutrinos, (b)~B-L local symmetry,
(c)~$SU(4)$-color, (d)~the SUSY unification scale, (e)~the seesaw
mechanism, and (f)~the pattern of $G(224)/SO(10)$ mass-matrices
allowed in the minimal Higgs system (see Sec.~4)-have played
crucial roles in realizing this \emph{unified and successful
description}. Before concluding, I now turn to a brief discussion
of proton decay in the next section.

\section{Proton Decay}

Perhaps the most dramatic prediction of grand unification is proton decay. I
have discussed proton decay in the context of the SUSY
$SO(10)/G(224)$-framework presented here in some detail in recent reviews
\cite{5DG(224),JCPErice} which are updates of the results obtained in
\cite{BPW1}. Here, I will present only the salient features
and the updated results. In SUSY unification there are in general three distinct mechanisms for proton decay.
\begin{enumerate}
\item \textbf{The familiar d=6 operators} mediated by $X$ and $Y$
gauge bosons of $SU(5)$ and $SO(10)$ As is well known, these lead
to $e^+\pi^0$ as the dominant mode with a lifetime $\approx
10^{35.3 \pm 1}$ yrs. \item \textbf{The ``standard'' $d=5$
operators} \cite{69SYW} which arise through the exchange of the
color-triplet Higgsinos which are in the $5_H +\bar{5}_H$ of
$SU(5)$ or $10_H$ of $SO(10)$. These operators require (for
consistency with proton lifetime limits) that the color-triplets
be made superheavy while the EW-doublets are kept light by a
suitable doublet-triplet splitting mechanism (for $SO(10)$, see
Ref.~\cite{DimWil,JCPEriceReview}. They lead to dominant
$\bar{\nu}K^+$ and comparable $\bar{\nu}\pi^+$ modes with
lifetimes varying from about $10^{29}$ to $10^{34}$ years,
depending upon a few factors, which include the nature of the
SUSY-spectrum and the matrix elements (see below).Some of the
original references on contributions of standard $d=5$ operators
to proton decay may  be found in
\cite{DimopRabyWilczek,Ellis,NathChemArno,Hisano,BabuBarr,BPW1,JCPErice,LucasRaby,Murayama}
\item \textbf{The so called ``new'' $d=5$ operators}
\cite{BPW2,JCPEriceReview} (see Fig.~1)\footnote{Note that in the
presence of a second SO(10)-singlet field $\bf{S'}$ carrying
flavor charge of 2a+1/2, an effective coupling of the form
$\bf{16_H.\overline{16_H}.S'}$ would be allowed preserving the
U(1)-flavor symmetry introduced in Sec. 4. With $\bf{S'}$ having a
VEV of GUT-scale, such a coupling would generate a mass-term
$\bf{16_H.\overline{16_H}}$ that enters into Fig. 1. The presence
of this second singlet $\bf{S'}$ does not in any way affect the
hierarchical pattern of effective couplings exhibited in Eqs. (9)
and (11). I thank Qaisar Shafi for raising this point.}
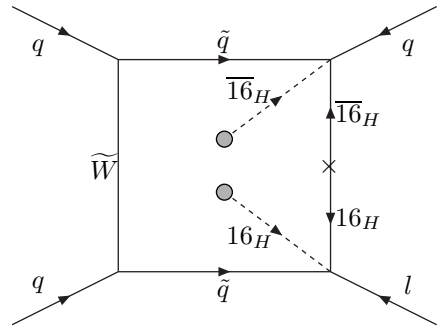
\begin{figure}
\begin{center}
\begin{picture}(200,140)(0,0)
 \ArrowLine(20,10)(60,30)
  \ArrowLine(20,130)(60,110)
   \ArrowLine(180,130)(140,110)
    \ArrowLine(180,10)(140,30)
\Text(30,25)[c]{$q$} \Text(30,115)[c]{$q$} \Text(170,115)[c]{$q$}
\Text(170,25)[c]{$l$}
 \ArrowLine(60,110)(140,110)
  \Line(60,30)(60,110)
  \ArrowLine(60,30)(140,30)
   \ArrowLine(140,70)(140,110)
    \ArrowLine(140,70)(140,30)
    \Text(140,70)[c]{$\times$}
 \Text(55,70)[c]{$\widetilde{W}$}
\Text(100,117)[c]{$\tilde{q}$} \Text(100,23)[c]{$\tilde{q}$}
\Text(151,50)[c]{$16_H$}\Text(151,90)[c]{$\overline{16}_H$}
\DashArrowLine(100,80)(140,110){2}
 \DashArrowLine(100,60)(140,30){2}
\GCirc(100,80){3}{0.7} \GCirc(100,60){3}{0.7}
\Text(110,99)[c]{$\overline{16}_H$}
 \Text(110,43)[c]{$16_H$}
\end{picture}
\caption{\label{Fig3}The ``new'' $d=5$ operators related to the
Majorana masses of the RH neutrinos. Note that the vertex at the
upper right utilizes the coupling in Eq.(11) which assigns
Majorana masses to $\nu_R$'s, while the lower right vertex
utilizes the $g_{ij}$ couplings in Eq.(9) which are needed to
generate CKM mixings.}

\end{center}
\end{figure}
which can generically arise through the exchange of color-triplet
Higgsinos in the  Higgs multiplets like $(16_H +\bar{16}_H)$ of $SO(10)$. Such
exchanges are possible by utilizing the joint effects of (a)~the couplings
given in Eq.~(\ref{eq:LMaj}) which assign superheavy Majorana masses to the RH
neutrinos through the VEV of $\bar{16}_H$, and (b)~the coupling of the form
$g_{ij} 16_i 16_j 16_H 16_H/M$ (see Eq.~(\ref{eq:Yuk})) which are needed, at
least for the minimal Higgs-system, to generate CKM-mixings. These operators
also lead to $\bar{\nu}K^+$ and $\bar{\nu}\pi^+$ as the dominant modes, and
they can quite plausibly lead to lifetimes in the range of
$10^{32}-10^{34}$ yrs [see below]. These operators, though most natural in a
theory with Majorana masses for the RH neutrinos, have been invariably omitted
in the literature.
\end{enumerate}

One distinguishing feature of the new $d=5$ operator is that they directly
link proton decay to neutrino masses via the Majorana masses of the RH
neutrinos. \emph{The other, and perhaps most important, is that these new
$d=5$ operators can induce proton decay even when the $d=6$ and standard $d=5$
operators mentioned above are absent.} This is what would happen if the string
theory or a higher dimensional GUT-theory would lead to an effective
$G(224)$-symmetry in $4D$ (along the lines discussed in Sec.~3), which would
be devoid of both $X$ and $Y$ gauge bosons and the dangerous color-triplets in
the $10_H$ of $SO(10)$. \emph{By the same token, for an effective
$G(224)$-theory, these new $d=5$ operators become the sole and viable source
of proton decay leading to lifetimes in an interesting range (see below).} And
this happens primarily because the RH neutrinos have a Majorana mass!

Our study of proton decay carried out in Ref.~\cite{BPW1} and
updated in \cite{5DG(224)} and \cite{JCPErice} has a few distinctive features:
(i)~It is based on a \emph{realistic framework for fermion masses and neutrino
oscillations}, as discussed in Sec.~4;(ii)~It includes the \emph{new $d=5$
operators} in addition to the standard $d=5$ and $d=6$ operators; (iii)~It
restricts \emph{GUT-scale threshold-corrections} to $\alpha_3(m_Z)$ so as to be
in accord with the demand of ``natural'' coupling unification and thereby
restricts $M_{eff}$ that controls the strength of the standard $d=5$ operators;
and (iv)~It allows for the ESSM extension \cite{BabuJi} of MSSM motivated on
several grounds (see e.g. \cite{BabuJi} and \cite{JCPErice}), which
introduces two vectorlike families in $16+\bar{16}$ of $SO(10)$ with masses of
order $1$ TeV, in addition to the three chiral families.

The study carried out in \cite{BPW1} and its updates, based on
recently reported values of the matrix elements $\beta_H$ and $\alpha_H$ and
the renormalization factors $A_L$ and $A_S$ for $d=5$ \cite{Turznyski}, and
$A_R$ for $d=6$ operators have been discussed
in detail in \cite{5DG(224)} and \cite{JCPErice}. In these reviews, I had used
the latest lattice-result available at the time which gave $\beta\approx 0.014$
GeV$^3$ \cite{Aoki}. This result was based, however, on quenching and finite
lattice spacing, which could introduce sizable systematic error. I had,
therefore, allowed an uncertainty by a factor of two either way and taken
$\beta_H = (0.014 \mbox{ GeV}^3)(1/2-2)$ and likewise
$\alpha_H = (0.015 \mbox{ GeV}^3)(1/2-2)$. A very recent calculation based on
quenched lattice QCD in the continuum limit yields:
$|\beta_H| = 0.0096(09)(^{+6}_{-20}) \mbox{ GeV}^3)$ and
$|\alpha_H| = 0.0090(09)(^{+5}_{-19}) \mbox{ GeV}^3)$ \cite{81Tsutsui}.
Allowing still for an uncertainty by $\sqrt{2}$ either way (due to quenching),
I now take $|\beta_H|$ and  $|\alpha_H| \gsim 0.0068$. This value however
nearly coincides with the value of $\beta_H$ and $\alpha_H$ at the lower end
$(\approx0.007 \mbox{ GeV}^3)$ used in previous estimates
\cite{5DG(224),JCPErice}. As a result \emph{the upper limits on proton lifetime
presented before remain practically unaltered}. The values of the parameters
now used are as follows:
$|\beta_H| \approx |\alpha_H|\approx(0.009\mbox{ GeV}^3)(1/\sqrt{2}-\sqrt{2})$;
$m_{\tilde{q}}\approx m_{\tilde{l}}\approx 1.2$ TeV $(1/2-2)$;
$(m_{\tilde{W}}/m_{\tilde{q}})=1/6(1/2-2)$;
$M_{H_C}(\mbox{min} SU(5))\leq 10^{16}$ GeV,
$A_L\approx 0.32$, $A_S\approx 0.93$,
$\tan{\beta}\leq 3$; $M_X\approx M_Y\approx 10^{16}$ GeV $(1\pm25\%)$, and
$A_R(d= 6, e^+\pi^0)\approx 3.4$.

The theoretical predictions for proton decay for the cases of minimal SUSY
$SU(5)$, SUSY $SO(10)$ and $G(224)$-models developed in Secs.~3 and 4, are
summarized in Table~2. They are obtained by following the procedure as in
\cite{JCPEriceReview,JCPErice} and using the parameters as mentioned
above.\footnote{The chiral Lagrangian parameter ($D+F$) and the
renormalization factor $A_{R}$ entering into the amplitude for
$p \rightarrow e^{+}\pi^{0}$ decay are taken to be 1.25 and 3.4 respectively.}
\begin{table*}
\centering
\begin{eqnarray}
\left. \frac{\mbox{SUSY $SU(5)$}}{\mbox{MSSM (std. $d=5$)}} \right\}
\begin{array}{l}\Gamma^{-1}(p \rightarrow \bar{\nu}K^{+}) \end{array} & \leq &
\begin{array}{lr}
\begin{array}{l} 1.2 \times 10^{31} \mbox{ yrs} \end{array} &
\left( \begin{array}{c} \mbox{Excluded by} \\ \mbox{SuperK} \end{array} \right)
\end{array} \label{new31}\\ \nonumber\\
\left. \frac{\mbox{SUSY $SO(10)$}}{\mbox{MSSM (std. $d=5$)}} \right\}
\begin{array}{l} \Gamma^{-1}(p \rightarrow \bar{\nu}K^{+}) \end{array} & \leq &
\begin{array}{lr}
\begin{array}{l} 1 \times 10^{33} \mbox{ yrs} \end{array} &
\left( \begin{array}{c} \mbox{Tightly constrained} \\
\mbox{by SuperK} \end{array} \right)
\end{array} \label{new32} \\ \nonumber\\
\left. \frac{\mbox{SUSY $SO(10)$}}{\mbox{ESSM (std. $d=5$)}} \right\}
\begin{array}{l}
\Gamma^{-1}(p \rightarrow \bar{\nu}K^{+})_{\mbox{Med.}} \\
\Gamma^{-1}(p \rightarrow \bar{\nu} K^{+})
\end{array}
&
\begin{array}{c} \approx \\ \lsim \end{array}
&
\begin{array}{lr}
\begin{array}{l}
(\mbox{1--10}) \times 10^{33} \mbox{ yrs} \\ 10^{35} \mbox{ yrs}
\end{array}
&
\left( \begin{array}{c} \mbox{Fully SuperK} \\ \mbox{Compatible}
\end{array} \right)
\end{array} \label{new33} \\ \nonumber\\
\left. \frac{\mbox{SUSY $G(224)/SO(10)$}}{\mbox{MSSM or ESSM (new $d=5$)}} \right\}
\begin{array}{l}
\Gamma^{-1}(p \rightarrow \bar{\nu}K^{+}) \\
B(p \rightarrow \mu^{+} K^{0})
\end{array}
&
\begin{array}{c} \lsim \\ \approx \end{array}
&
\begin{array}{lr}
\begin{array}{l}
2 \times 10^{34} \mbox{ yrs} \\ (1-50)\%
\end{array}
&
\left( \begin{array}{c} \mbox{Fully Compatible} \\ \mbox{with SuperK}
\end{array} \right)
\end{array} \label{new34} \\ \nonumber\\
\left. \frac{\mbox{SUSY $SU(5)$ or $SO(10)$}}{\mbox{MSSM ($d=6$)}} \right\}
\begin{array}{l} \Gamma^{-1}(p \rightarrow e^{+} \pi^{0}) \end{array} & \approx
& \begin{array}{lr} \begin{array}{l} 10^{35 \pm 1} \mbox{ yrs}
\end{array} & \left( \begin{array}{c} \mbox{Fully Compatible} \\
\mbox{with SuperK}
\end{array} \right)
\end{array} \label{new35}
\end{eqnarray}
\caption{A Summary of Results on Proton Decay}
\end{table*}

It should be stressed that the upper limits on proton lifetimes
given in Table~2 are quite conservative in that they are obtained
(especially for the top two cases) by stretching the uncertainties
in the matrix element and the SUSY spectra
to their extremes so as to prolong proton
lifetimes.  In reality, the lifetimes should be shorter than the
upper limits quoted above.

Now the experimental limits set by SuperK studies are as follows
\cite{SKlimit}:
\begin{eqnarray}
\Gamma^{-1}(p\rightarrow e^{+}\pi^{0})_\mathrm{expt} & \geq &
    6 \times 10^{33}\mbox{ yrs} \nonumber \\
\Gamma^{-1}(p\rightarrow \bar{\nu}K^{+})_{\mathrm{expt}} & \geq &
    1.6 \times 10^{33}\mbox{ yrs}
\end{eqnarray}
The following comments are in order.
\begin{enumerate}
\item By comparing the upper limit given in Eq.~(\ref{new31}) with
the experimental lower limit, we see that the
\emph{minimal} SUSY $SU(5)$ with the conventional MSSM spectrum is
clearly excluded by a large margin by proton decay searches.  This
is in full agreement with the conclusion reached by other authors
(see e.g. Ref.~\cite{Murayama}).\footnote{See, however,
Refs.~\cite{BajcPerezSenja} and \cite{EmmanuelWies}, where attempts are made
to save minimal SUSY SU(5) by a set of scenarios.
These include a judicious choice of sfermion mixings, higher
dimensional operators and squarks of first two families having masses of
order 10 TeV.}
\item By comparing
Eq.~(\ref{new32}) with the empirical lower limit,
we see that the case of MSSM embedded in
$SO(10)$ is already tightly constrained to the point of being
disfavored by the limit on proton lifetime.  The constraint is of
course augmented by our requirement of \emph{natural coupling
unification}, which prohibits accidental large cancelation
between different threshold corrections (see \cite{BPW1}).
\item In
contrast to the case of MSSM, that of ESSM \cite{BabuJi} embedded in $SO(10)$
(see Eq.~(\ref{new33})) is fully compatible with the SuperK limit.
In this case, \(\Gamma_{\mathrm{Med}}^{-1}(p \rightarrow
\bar{\nu}K^{+}) \approx 10^{33} - 10^{34} \mbox{ yrs}\), given in
Eq.~(\ref{new33}), corresponds to the parameters involving the
SUSY spectrum and the matrix element $\beta_{H}$ being in the
\emph{median range}, close to their central values.
\item We see from Eq.~(\ref{new34}) that
the contribution of the new operators related to the Majorana
masses of the RH neutrinos (Fig.~1) (which is the same for MSSM
and ESSM and is independent of $\tan \beta$) is fully compatible
with the SuperK limit.  These operators can quite naturally lead
to proton lifetimes in the range of $10^{33}-10^{34}$ yrs with an
upper limit of about \(2 \times 10^{34}\) yrs.
\end{enumerate}

In summary for this section, within the $SO(10)/G(224)$ framework and with the
inclusion of the standard as well as the new $d=5$ operators, one obtains
(see Eqs.~(31)--(35)) a conservative upper limit on proton lifetime given
by:
\begin{equation}
\begin{array}{rl}
\tau_{\mathrm{proton}} \lsim (1/3 - 2) \lefteqn{\times 10^{34} \mbox{ yrs}}
 & \nonumber \\
 & \left( \begin{array}{c} \mbox{SUSY} \\ SO(10)/G(224) \end{array} \right)
\end{array}
\label{new36}
\end{equation}
with $\bar{\nu}K^{+}$ and $\bar{\nu}\pi^{+}$ being the dominant modes and
quite possibly $\mu^{+}K^{0}$ being prominent.

The $e^+\pi^0$-mode induced by gauge boson-exchanges should have an
inverse decay rate in the range of $10^{34}-10^{36}$ years (see Eq.
(35)).  The implication of these predictions for a next-generation detector
is noted in the next section.

\section{Concluding Remarks}

The neutrinos seem to be as elusive as revealing. Simply by virtue
of their tiny masses, they provide crucial information on the
unification-scale, and even more important on the nature of the
unification-symmetry. In particular, as argued in Secs.~4 and 6,
(a)~the magnitude of the superK-value of $\sqrt{\delta m^2_{23}}
(\approx 1/20 \mbox{ eV})$, (b)~the $b/\tau$ mass-ratio, and
(c)~baryogenesis via leptogenesis, together, provide clear support
for (i)~the existence of the RH neutrinos, (ii) the existence of
B-L as a local symmetry in 4D above the GUT-scale, (iii)~the
notion of $SU(4)$-color symmetry which provides not only the RH
neutrinos but also B-L as a local symmetry and a value for
$m(\nu^{\tau}_{\mbox{Dirac}})$, (iv)~the familiar SUSY
unification-scale which provides the scale of $M_R$, and last but
not least, (v)~the seesaw mechanism. \emph{In turn this chain of
argument selects out the effective symmetry in $4D$ being either a
string-derived $G(224)$ or $SO(10)$-symmetry, as opposed to the
other alternatives like $SU(5)$ or flipped $SU(5)'\times U(1)$}.

It is furthermore remarkable that the tiny neutrino-masses also seem to hold
the key to the origin of baryon excess and thus to our own origin!

In this talk, I have tried to highlight that the $G(224)/SO(10)$-framework as
described here is capable of providing a \emph{unified description} of fermion
masses, neutrino oscillations, CP and flavor violations as well as of
baryogenesis via leptogenesis. This seems non-trivial. The framework
is also highly predictive and can be further tested by studies of CP and flavor
violations in processes such as (a)~$B_d\rightarrow\phi K_S$-decay,
(b)~$(B_S,\bar{B}_S)$-decays, (c)~edm of neutron, and (d)~leptonic flavor
violations as in $\mu\rightarrow e\gamma$ and
$\tau\rightarrow\mu\gamma$-decays.

To conclude, the evidence in favor of supersymmetric grand unification, based
on a string-derived $G(224)$-symmetry (as described in Sec.~3) or
$SO(10)$-symmetry, appears to be strong. It includes:
\begin{itemize}
\item Quantum numbers of all members in a family, \item
Quantization of electric charge, \item Gauge coupling unification,
\item $m^0_b\approx m^0_\tau$ \item $\sqrt{\delta
m^2(\nu_2-\nu_3)}\approx 1/20$ eV, \item A maximal
$\Theta^{\nu}_{23}\approx \pi/4$ with a minimal $V_{cb} \approx
0.04$, and \item Baryon Excess $Y_B\approx 10^{-10}$.
\end{itemize}
All of these features and more including (even) CP and flavor
violations hang together neatly within a single unified framework
based on a presumed string-derived four-dimensional $G(224)$ or
$SO(10)$-symmetry, with supersymmetry. It is hard to believe that
this neat fitting of all these pieces can be a mere coincidence.
It thus seems pressing that dedicated searches be made for the two
missing pieces of this picture-that is supersymmetry and proton
decay. The search for supersymmetry at the LHC and possible future
NLC is eagerly awaited. That for proton decay will need a
next-generation megaton-size underground detector.

\textbf{Acknowledgments}: I would like to thank Kenzo Nakamura and Yoji
Totsuka for their kind hospitality. I have benefited from many collaborative
discussions with Kaladi S. Babu, Parul Rastogi and Frank Wilczek on topics
covered in this lecture.  I would also like to thank Keith Olive,
Stuart Raby, Goran Senjanovic, Qaisar Shafi
and N. Tsutsui for most helpful communications, and Antonio Masiero for
very helpful comments.
The research presented here is
supported in part by DOE grant No.~DE-FG02-96ER-41015.

\bibliographystyle{unsrt}
\bibliography{pati}

\begin{thebibliography}{10}
\bibitem{FN26}
For G(224), one can choose the corresponding sub-multiplets -- that is (1, 1,
  15)$_H$, (1, 2, $\bar{4}$)$_H$, (1, 2, 4)$_H$, (2, 2, 1)$_H$ -- together with
  a singlet $S$, and write a superpotential analogous to Eq. (\ref{eq:Yuk}).
\bibitem{newFN36}
Note that the magnitudes of $\eta$, $\epsilon$ and $\sigma$ are fixed by the
  input quark masses. Furthermore, one can argue that the two contributions for
  $\theta_{\nu_\mu\nu_\tau}^{\rm osc}$ [see Eq. (\ref{eq:pred})] necessarily
  add to each other as long as $|y|$ is hierarchical ($\sim 1/10$) \cite{BPW1}.
  As a result, once the sign of $\epsilon$ relative to $\eta$ and $\sigma$ is
  chosen to be negative, the actual magnitudes of $V_{cb}\approx (0.044)$ and
  $\sin^2 2\theta_{\nu_\mu\nu_\tau}^{\rm osc}\approx 0.92-0.99$ emerge as
  predictions of the model \cite{BPW1}.
\bibitem{FN32}
Note that such an operator would be allowed by the flavor symmetry defined in
  Eq.~(10) if one sets $a=1/2$. In this case, operators such as $W_{23}$ and
  $W_{33}$ that would contribute to $\nu_L^\mu\nu^\tau_L$ and
  $\nu_L^\tau\nu^\tau_L$ masses would be suppressed relative to $W_{12}$ by
  flavor symmetry.
\end{thebibliography}
\end{document}